\documentclass[useAMS,usenatbib]{mn2e}
\usepackage{graphicx}
\usepackage{latexsym}

\title[The structure of molecular clouds and the universality of the clump mass function]{The structure of molecular clouds and the universality of the clump mass function}

\author[Smith, Clark \& Bonnell]{Rowan J. Smith$^{1}$ \thanks{Email: rjs22@st-andrews.ac.uk}, Paul C. Clark$^{2}$ and Ian A. Bonnell$^{1}$ \\
$^1$ SUPA, School of Physics \& Astronomy, University of St Andrews, North Haugh, St Andrews, Fife, KY16 9SS, UK \\
$^2$ Zentrum f\"ur Astronomie der Universit\"at Heidelberg, Institut f\"ur Theoretische Astrophysik,Albert-Ueberle-Str. 2, Heidelberg, Germany  
}

\begin{document}

\pagerange{\pageref{firstpage}--\pageref{lastpage}} \pubyear{2008}

\maketitle

\label{firstpage}

\def\solmas{{M$_\odot$}}
\def\solm{{M_\odot}}
\def\mnras{MNRAS}
\def\apj{ApJ}
\def\aap{A\&A}
\def\apjl{ApJL}
\def\apjs{ApJS}
\def\bain{BAIN}
\def\pasp{PASP}
\def\araa{ARA\&A}
\def\ga{\sim}

\begin{abstract}

Using an SPH simulation of a star-forming region in a molecular cloud, we show that the emergence of a clump mass function (CMF) resembling the stellar initial mass function (IMF) is a ubiquitous feature of molecular cloud structure, but caution against its over-interpretation. We employ three different techniques to extract the clumps in this study. In the first two, we interpolate the SPH particle data to 2 and 3 dimensional grids before performing the clump-find, using position-position (PP) and position-position-velocity (PPV) information respectively. In the last technique, the clump-finding is performed on the SPH data directly, making use of the full 3 dimensional position information. Although the CMF is typically similar to that observed in regions of nearby star formation, the individual clumps and their masses are found to be unreliable since they depend strongly on the parameters and the method of the clump-finding. In particular we find that the resolution and orientation of the data make a significant difference to the resulting properties of the identified clumps in the PP and PPV cases. We conclude that making comparisons between a CMF and the stellar IMF should be done with caution, since the definition of a clump boundary, and hence the number of clumps and their properties,  
are arbitrary in the extraction method. This is especially true if molecular clouds are truly scale free.

\end{abstract}


\section{Introduction}
\label{sec:introduction}

Recent observations \citep{Motte98,Johnstone00,Johnstone06, Nutter07,Alves07,Ward-Thompson07} of dense gas cores in molecular clouds have shown their distribution resembles the stellar initial mass function (IMF). It is hoped that by studying these dense cores  the earliest stages of star formation can be probed and the origins of the IMF revealed.

\citet{Motte98} have shown that the clump mass function (CMF) in $\rho$ Oph can be described by a similar power-law fit as the IMF. Above $\sim 0.5$ M$_{\odot}$ they find $dn/dm \propto m^{-\alpha}$ is well-fitted by an $\alpha = 2.5$, while at lower masses they see a turn-over that can be fitted with $\alpha = 1.5$. These values for $\alpha$ are broadly consistent with the usual fits to the IMF \citep{Kroupa02}. Similar results have been confirmed by a number of authors for a variety of nearby star-forming regions, although the range of core masses found and the break mass of the power law fit can vary \citep{Johnstone00,Johnstone06, Nutter07,Alves07, Testi98}. A more detailed account of the clump/core mass distribution can be found in \citet{Ward-Thompson07}.

One interpretation of the similarity between the CMF and IMF is that the masses of stars are controlled by the fragmentation of the cloud. \citet{Myers08} propose that for cores with a high density contrast to their surroundings, the resulting protostellar mass will be the gas mass whose free-fall time equals the core dispersal time due to outflows. Additionally \citet{Swift08} take a slightly different approach and argue that the resulting shape of an IMF formed from a Saltpeter-like CMF is robust against different core evolution scenarios. 

In this paper, we show that the shape of clump mass function seems to be a ubiquitous feature of molecular clouds, but the identified clumps and their masses are sensitive to the method used to find them.

Molecular cloud (MC) structure as a whole is complex, filamentary and extends through all observable scales (\citealt{Williams00}). It can also be thought of as hierarchical: small dense cores are part of small clumps of gas which are themselves part of larger gas clumps. This continues over the full spatial range of observations \citep{Bergin07}. MCs also exhibit supersonic turbulence. \citet{Larson81} showed that the internal velocity dispersion $\sigma(v)$ of a cloud is related to its size $L$ by the relation $\sigma(v)\propto L^{-0.38}$. Such a scale-free relation is a natural outcome of a turbulent cascade, with \citet{Brunt04} proposing that compressible shock dominated Burgers turbulence is most consistent with observations.

As supersonic flows compress the gas in the molecular cloud they create a complex filamentary density field, suggesting that the observed structure in MCs is a direct consequence of turbulent motions \citep{Klessen00, Klessen01}. At the smallest scales of the hierarchy, the transonic dense cores formed by the turbulent cascade may be the precursors of star formation (\citealt{Goodman98}, \citealt{Ward-Thompson07}). Indeed \citet{Padoan02} have predicted that mass distribution of Jeans unstable clumps created by super-Alvenic turbulence will resemble the IMF, a prediction that is supported by the observations of the CMFs discussed above.  

However there are still questions as to whether the CMF observations can unambiguously support the cloud fragmentation picture. Typically, the observations must rely on two-dimensional column density maps (PP), which are limited by resolution and sensitivity, and the observed features in the column density may be contaminated by integrated emission along the line-of-sight. Three-dimensional information is only available by obtaining position-position-velocity (PPV) cubes of the emission from a particular molecular tracer. Converting between emission and column density relies on many assumptions \citep{Pineda08} and accounting for the properties of dust is itself challenging \citep{Schnee06}. Further, the observed emission is itself dependent on the entangled effects of density, temperature, molecular tracer properties, and the observational technique \citep{DiFrancesco07}.

In this paper, we investigate how the differences in the form of the data -- whether it be PP, PPV or full three-dimensional PPP -- affects the resulting CMF. We explore how the resolution the data and orientation of the cloud affect the mass functions profile. Since we are interested in comparing observational style clump-finds to a full three dimensional clump-find, the data must necessarily be synthetic. As such we use the results from a high-resolution smoothed particle hydrodynamics (SPH) simulation of driven turbulence. A full description of this simulation can be found in Section \ref{sec:simulation}. To produce the PP and PPV style clump-finds the SPH data is interpolated to a grid, while for the PPP data the clump-find is done directly on the SPH data. The method used to extract the clumps is similar to that described by \citet{Williams94}, although there are some differences that can be found in Section \ref{sec:clumpfind}.

\section{The Simulation}
\label{sec:simulation}

We use a three dimensional smooth particle hydrodynamics (SPH) code \citep{Monaghan92}, to simulate star formation in a collapsing gas cloud. The SPH code has variable smoothing lengths and artificial viscosity \citep{Gingold83}. Self-gravity is calculated via a binary tree \citep{Benz90} and sink particles \citep{Bate95} are used to model the star formation. Table \ref{simprops} shows the initial conditions of the simulation. The molecular cloud has a mass of $118.2$ M$_{\odot}$, roughly comparable to the masses of the `cores' in $\rho$ Oph \citep{Motte98}. The cloud is modelled with $\sim$ 5 million SPH particles, which gives a mass resolution of $2.36 \times 10^{-3}$ M$_{\odot}$ \citep{Bate97}. As such, we are able to resolve clumps with masses in the brown dwarf regime. The simulation was run on the SUPA Altix computer at The University of St Andrews.

\begin{table}
	\centering
		\caption{All the clump-finding presented in this paper were performed on a single SPH simulation. The initial conditions of our simulation are given here. The mass resolution is the minimum mass gravitational forces can be resolved for and is calculated via $M_{res} \sim 100 M_{total}/N_{part}$. The Jeans mass, free-fall time and sound speed are calculated from the average density of the simulation before collapse using the formulae $M_{J}=(4\pi \rho/3)^{-1/2} (5kT/(2G\mu))^{3/2}$, $t_{ff}=(3\pi/(32G\rho))^{1/2}$. }
		\begin{tabular}{l c }
   	         \hline
	         \hline
	         Size & $2$ pc cube\\
	         Mass & $118.2$ M$_{\odot}$\\
	         Particles & $5000211$\\
	         Mass resolution & $2.36 \times 10^{-3}$ M$_{\odot}$\\
	         Jeans mass & $ 5.7$ M$_{\odot}$\\
	         Free-fall time & $ 2.1\times 10^{6}$ yrs\\
	         Sound speed & $0.18$ kms$^{-1}$\\
		\hline
		\end{tabular}
	\label{simprops}
\end{table}

We ensure the velocity and density fields of the simulation are self-consistent as follows. A turbulent velocity grid, consistent with a Larson velocity dispersion of $\sigma \propto r^{0.5}$,  is generated according to \citet{Dubinski95} and \citet{Myers99}. The SPH particle velocities are interpolated from this grid. The grid is continually re-imposed on the particles as they are allowed to relax without self-gravity. The driving pattern is not fixed, but rather evolves between two realisations on a crossing time.  After half a free fall time the probability distribution function has a lognormal shape as is expected for an isothermal gas without self-gravity, \citep{Klessen00}, and the turbulence has power on all scales. The turbulence has a r.m.s velocity of Mach 4.25 which corresponds to the simulated gas cloud being just kinetically supported (E$_{k}$=$|$E$_{p}|$).

\begin{figure}
\begin{center}
\includegraphics[width=3in]{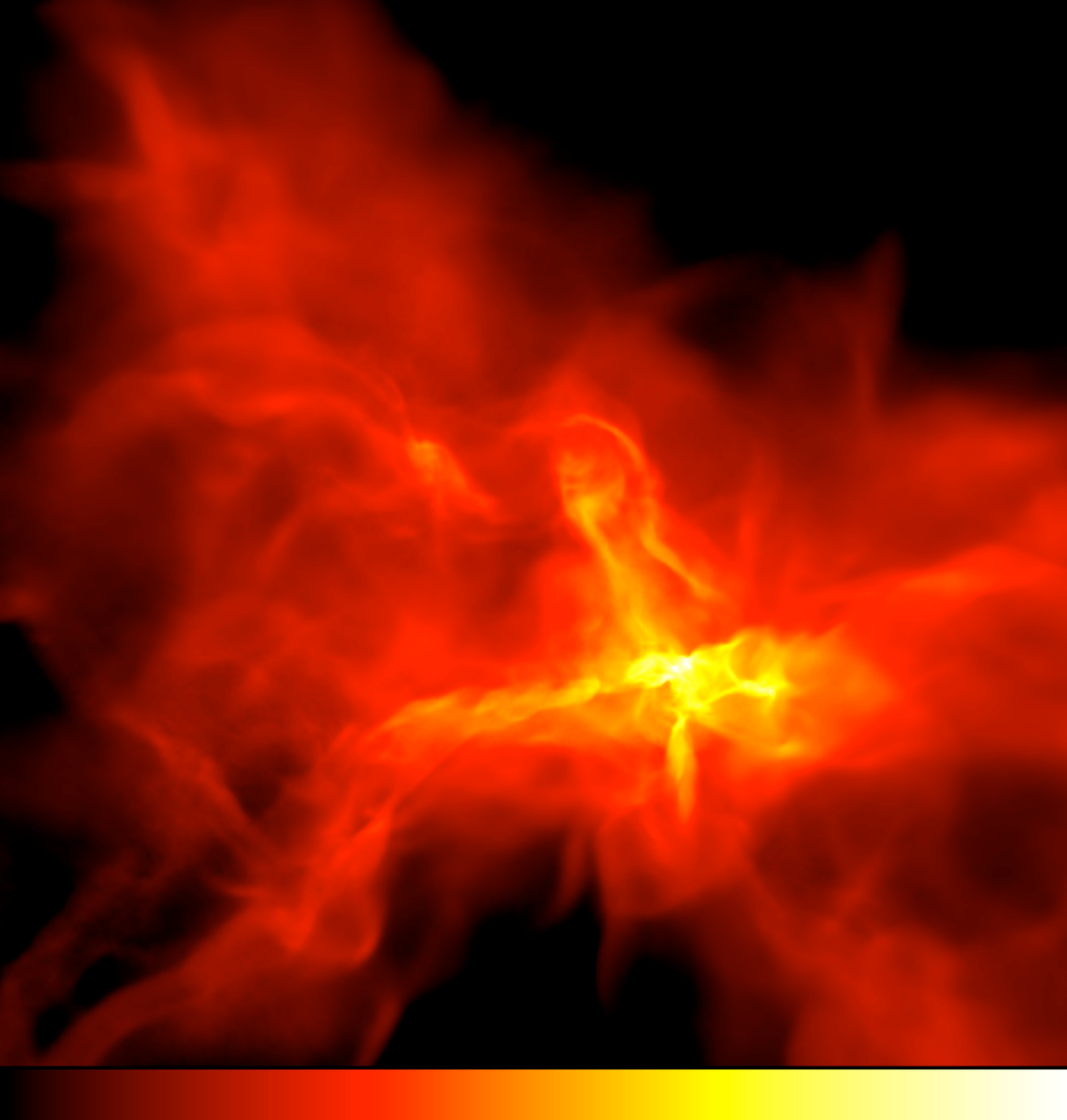}
\caption{The column density of the data file used for the clumpfinding comparison. The density scale is logarithmic and runs from $0.005$ gcm$^{-2}$ (black) to $1$ gcm$^{-2}$ (white). The region shown is $1$ pc by $1$ pc large and contains $79.5$ M$_{\odot}$.}
\label{standard}
\end{center}
\end{figure}

When self-gravity is switched on, we stop the turbulent driving, allowing the supersonic motions to decay. \citet{MacLow98} have shown that turbulence typically decays on a crossing time, however, the turbulence in our simulation remains supersonic throughout due to the conversion of potential energy to kinetic energy as the cloud contracts. Star formation, as modelled by the formation of the first sink particle, occurs after one free-fall time $2.1\times 10^{6}$ yrs. At this point the molecular cloud has the filamentary structure typically seen in observations (e.g. \citealt{Bergin07}). We do not follow the evolution of the cloud beyond the formation of the first sink (that is, the runaway collapse of the first core). Rather, we use a snapshot of the cloud structure immediately before the formation of the first sink, as the input data for our clump-finding comparison. Figure \ref{standard} shows a column density projection of the cloud at this point in the evolution.

A polytropic equation of state is used to mimic the evolution between the optically thin and optically thick regimes. Below densities of $10^{-13}$ gcm$^{-3}$ the value of the exponent gamma is initially set to $\gamma=1$ to mimic dust cooling. At gas densities of $10^{-13}$ gcm$^{-3}$ it changes to $\gamma=7/5$ where the gas becomes optically thick. Finally at densities greater than $10^{-7}$ gcm$^{-3}$ it is set back to $\gamma=1$ to model the dissociation of H$_{2}$. It is during this last isothermal regime that sinks are allowed to form.

\section{ Clumpfinding Methods}
\label{sec:clumpfind}

First, we wish to stress the nomenclature used in this paper. \citet{Williams00} were among the first to try make a clear definition for the terms `clump' and `core' and \citet{Bergin07} have refined this. They define a clump as having a size $0.3-3$ pc and containing $50-500$ M$_{\odot}$ and define a core as being around $0.03-0.2$ pc and having a mass of $0.5-5$ M$_{\odot}$. For this paper we shall simply use the term clump to refer to any object that is overdense compared to its surroundings and has been identified by our clump-finding algorithm (that is, appears when the cloud is viewed in the chosen contours). By the above definition however, most of our objects are technically cores.

The basic approach of the clump-finding algorithm is using a 'friend-of-friends' algorithm start from the point with the highest density, and move to lower densities, assigning the cells/particles to the same clumps if they are neighbours. The first position will be assigned as the head cell of a clump, if the next highest emitting region is a neighbour of the previous one then it is assigned to the same clump, if not it is the head cell of a new clump. Contour levels are used to determine whether a clump is a significant feature or merely a bump on a larger clump. A similar technique has been used by other authors (e.g. \citealt{Klessen00}). If the heads of two neighbouring clumps and their intersection are on the same density contour the clumps are merged. If a clump shares neighbouring cells with another but its head is on a lower density contour then the smaller clump is merged with the larger one. This process continues until all the points have been considered. Although this method of clump-finding is similar to the publicly available CLUMPFIND algorithm \citep{Williams94}, we do include a modification: the clumps are defined down to either a minimum density, or the lowest contour which it shares with a neighbouring clump. As such the mass in the clumps is typically much less than the mass in the cloud. By doing this, we are able to mimic the background subtraction that is normally performed on observational data sets. The definition of a `clump' here is also more in keeping with those studied by \citet{Motte98} and  \citet{Padoan07}.

As Pineda et al (in preparation) have shown, clump-finding routines are very sensitive to the positions and number of contour levels.  In this study we ensured that the contour levels were static in all the cases considered here. The lower density sensitivity was usually set at a column density of $0.02$ gcm$^{-2}$, we investigate the effect of varying minimum density in Section \ref{subsec:den}. 

In this study, we apply our clump finding algorithm to three different styles of data set: position-position (PP), position-position-velocity (PPV) and position-position-position (PPP). The first two (PP and PPV) represent the type of data that one can obtain from molecular cloud surveys. In these cases the `density' on which the clump-find is performed is the column density projected along the line of sight. In the PPV case, one simply has the projected column density per velocity channel. We obtain these maps by interpolating the SPH particles to a grid of given resolution. Rather than interpolating to a full PPP grid and then integrating along the required line of sight, it is more efficient to integrate the spherically symmetric SPH smoothing kernel along the line of sight, and use the resulting 2D kernel to perform the interpolation. The full PPP clump-find is performed directly on the SPH particle data, although the process of clump-finding is exactly the same.

\section{Two-dimensional (PP) Clumpfinding}
\label{sec:pp}

As mentioned in Section \ref{sec:introduction}, there have been a number of attempts to connect the structure of the gas in dense star-forming regions to the IMF. The majority of the work to date has produced results based on 2D maps of these regions, either using dust continuum emission (e.g \citealt{Johnstone06}) or extinction mapping (e.g \citealt{Lada08}) to obtain the column density distribution in the plane of the sky. As such, we start our clump-finding analysis on position-position column density maps of our simulated molecular cloud. 

\begin{figure}
\begin{center}
\includegraphics[width=3in]{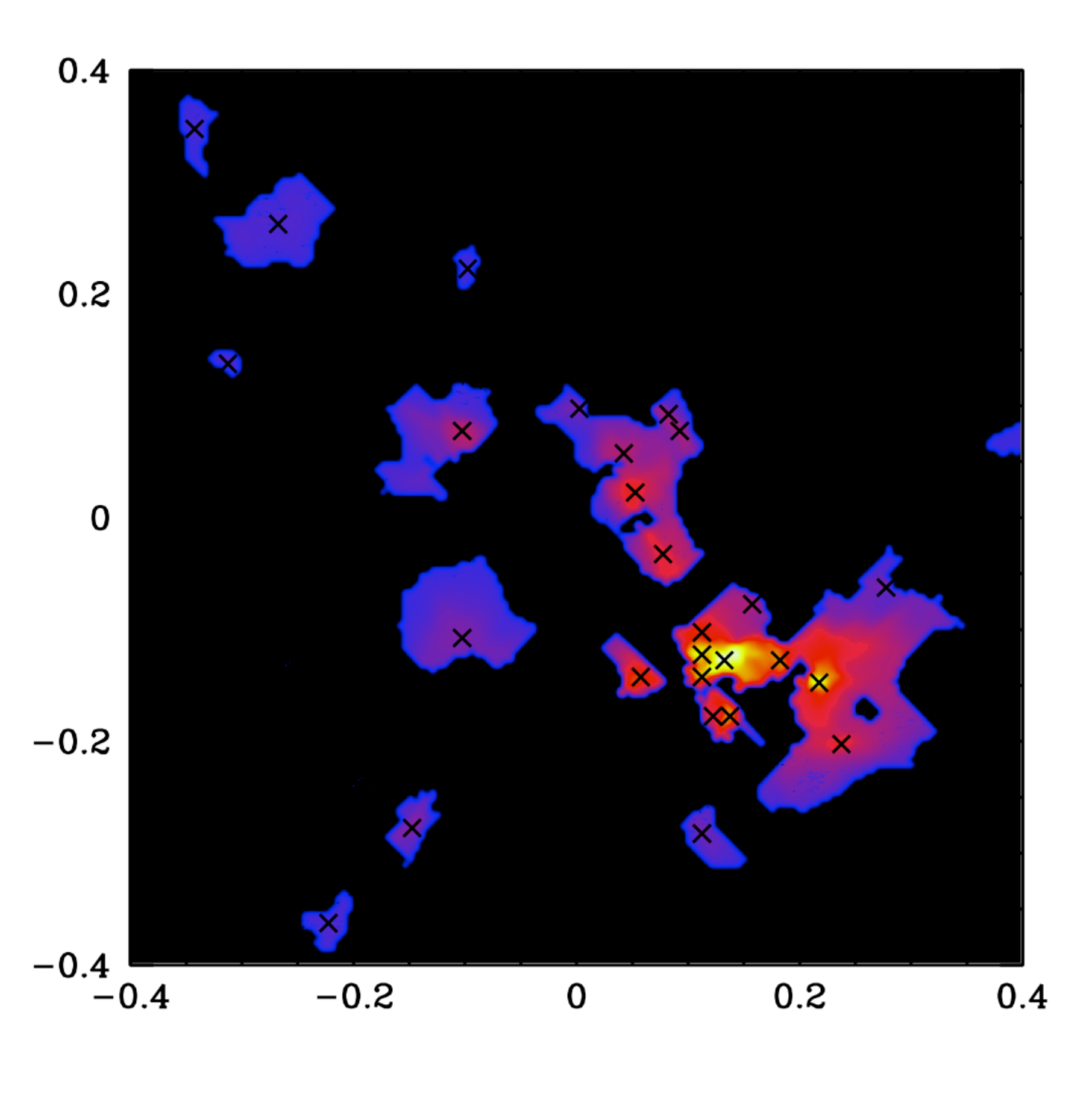}
\caption{The standard PP clump-find projected in the xy plane, crosses denote the centre of the clumps. This clump-find used data from a $200 \times 200$ grid.  The scale is shown in parsecs and the colours represent column densities in the range $0.02$ gcm$^{-2}$ (blue) to $1$ gcm$^{-2}$ (yellow) at logarithmic intervals.}
\label{sclumpfind}
\end{center}
\end{figure}

Figure \ref{sclumpfind} shows the PP clumpfind we adopt as the standard which we make comparisons against. The density is shown as a column density projection of the clumps after they have been background subtracted, the same scale is used as in Figure \ref{standard} for ease of comparison and to allow the clumps to be seen clearly, though no clump has a density lower than the minimum value of $0.02$ gcm$^{-2}$. In the regions where no clump has been assigned the lowest colour on the density scale is used. Table \ref{Parameters} shows the parameters used to obtain the PP standard clumpfind.

\begin{table}
	\centering
		\caption{The parameters used to find our standard PP clumpfind shown in Figure \ref{sclumpfind}. The resolutions grid cells correspond to a physical length of $1,031$ AU.}
		\begin{tabular}{l c}
   	         \hline
	         \hline
	         Minimum Density & $0.02$ $gcm^{-2}$ \\
	         Projection & xy plane \\
	         Resolution & $200\times200$ grid cells\\
	         Spatial Resolution &  $1,031$ AU \\
	         Density Countours & $5$ per magnitude\\
		\hline
		\end{tabular}
	\label{Parameters}
\end{table}

\subsection{Resolution}
\label{subsec:res}
\begin{figure*}
\begin{center}
\begin{tabular}{c c }
\includegraphics[width=3in]{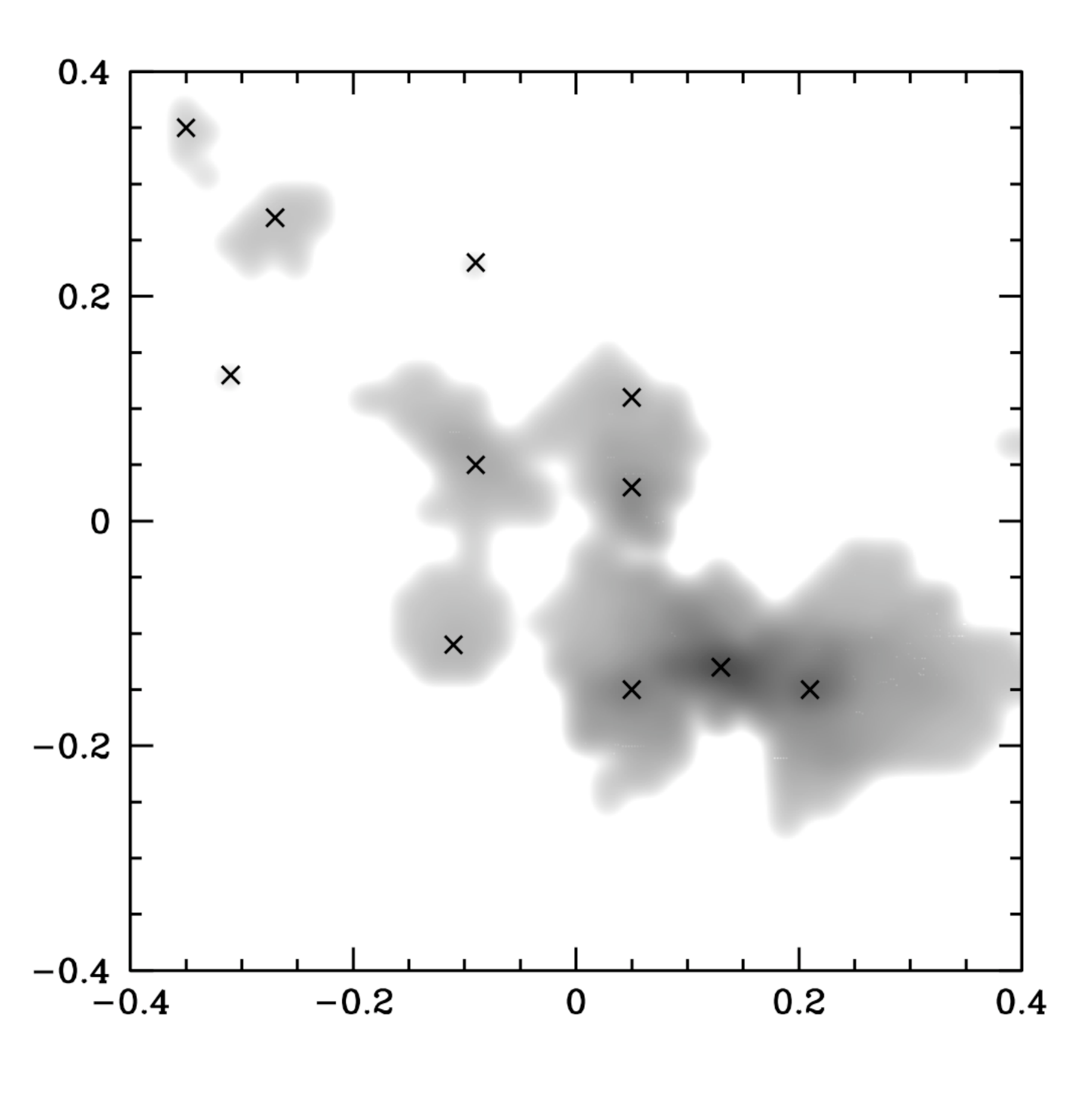} &
\includegraphics[width=3in]{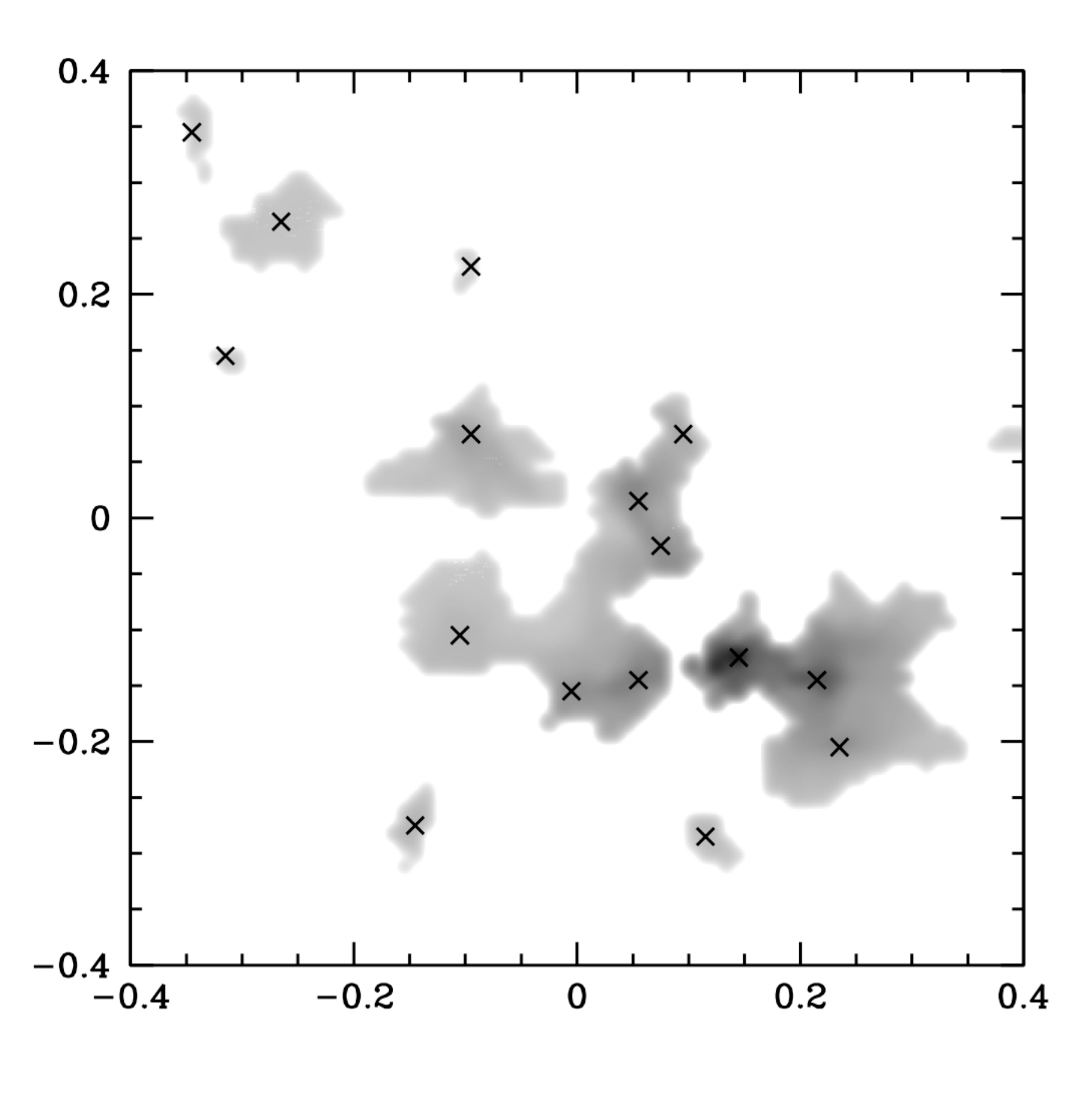} \\
\includegraphics[width=3in]{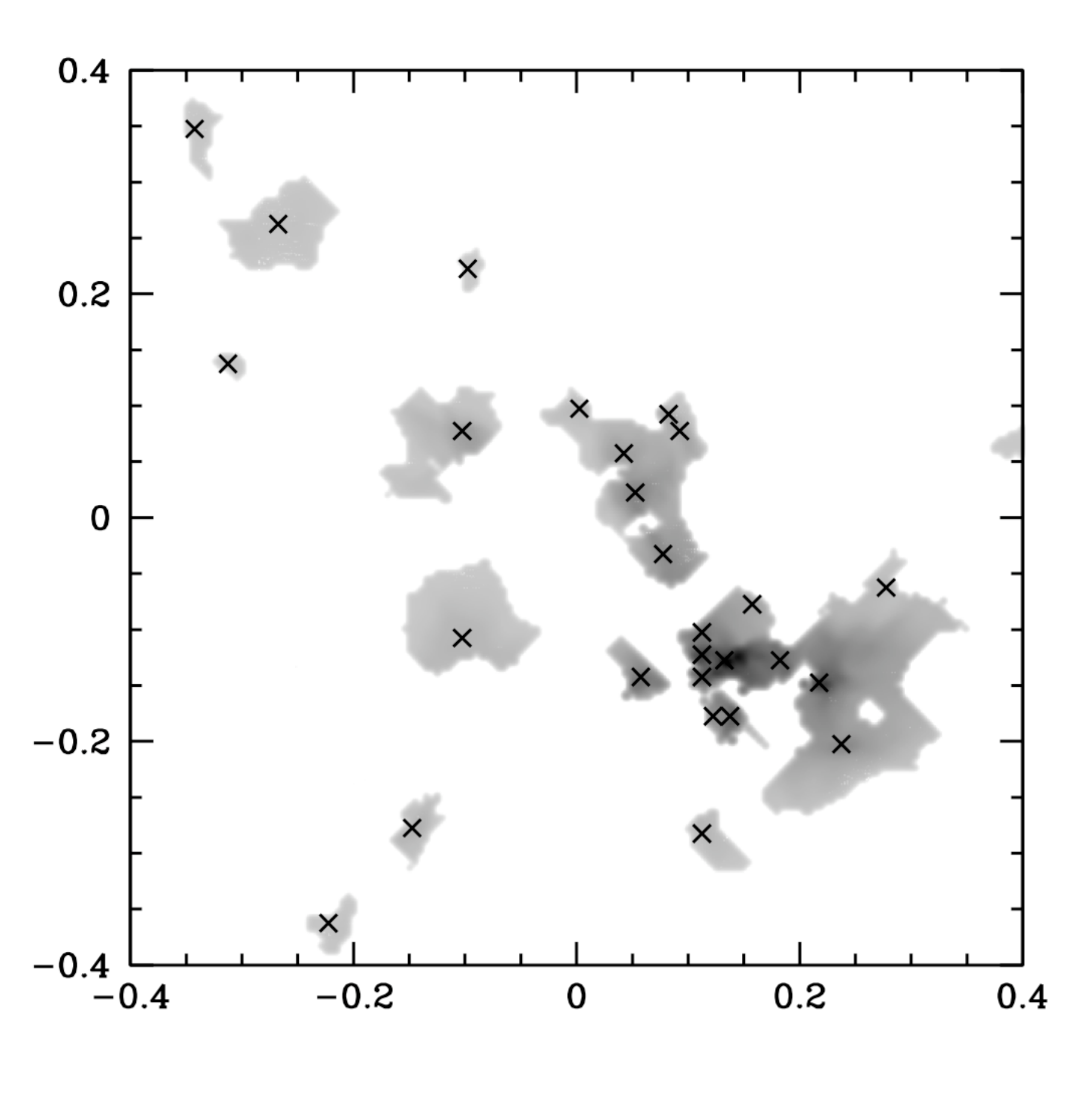} &
\includegraphics[width=3in]{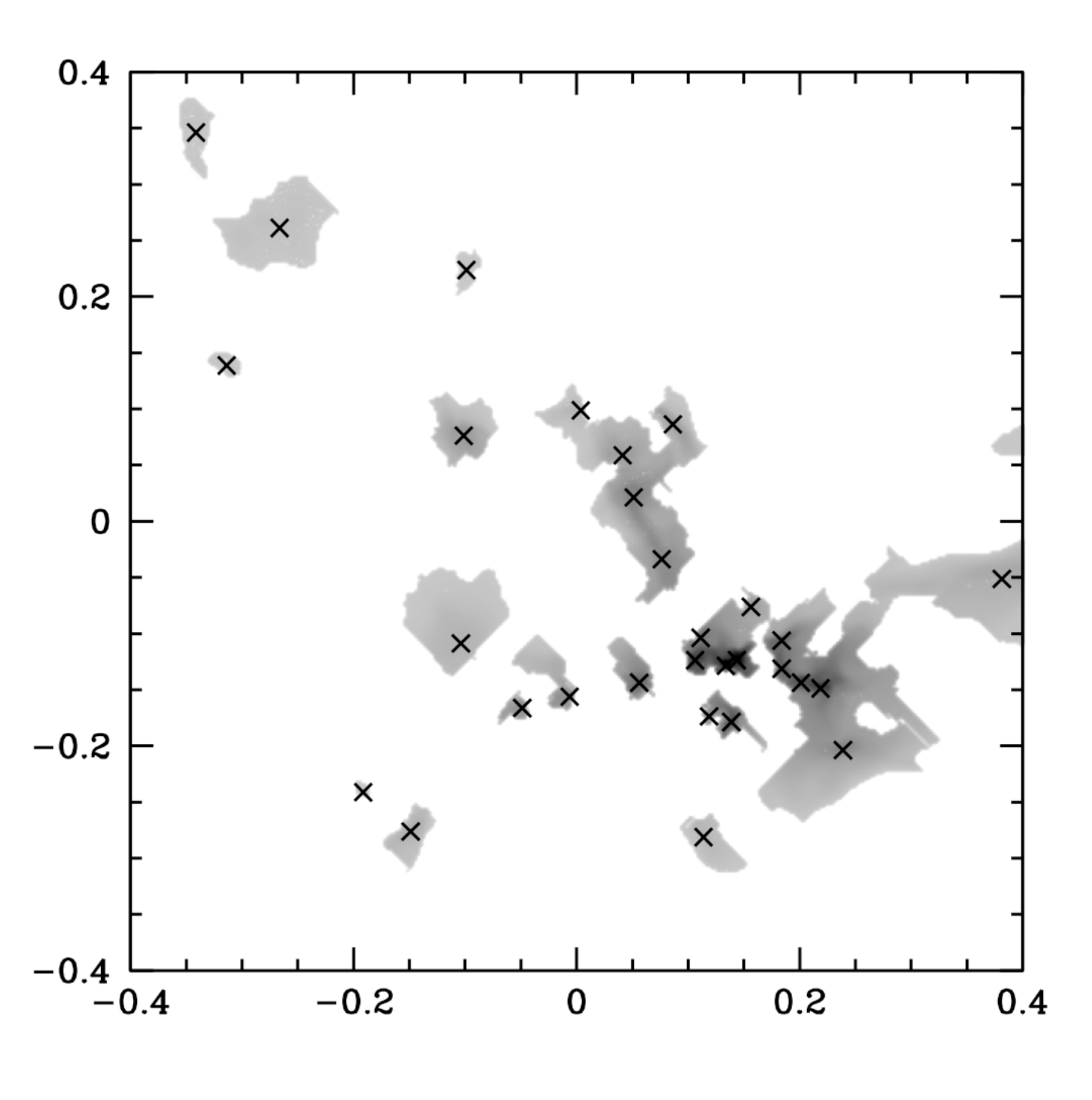} \\
\end{tabular}
\caption{Four clumpfinds carried out on the central region where star formation will occur with varying resolution levels. \textit{Top left} A $50\times50$ grid with a cell width of $4,125$ AU, \textit{top right} a $100\times100$ grid with a cell width of $2,063$ AU, \textit{bottom left} a $200\times200$ grid with a cell width of $1,031$ AU and \textit{bottom right} a $400\times400$ grid with a cell width of $516$ AU. The scale is shown in parsecs and the grayscale represent column densities in the range $0.02$ gcm$^{-2}$ (grey) to $1$ gcm$^{-2}$ (black) at logarithmic intervals. Crosses show the centre of the clumps.}
\label{Resolution}
\end{center}
\end{figure*}

\begin{figure}
\begin{center}
\includegraphics[width=3in]{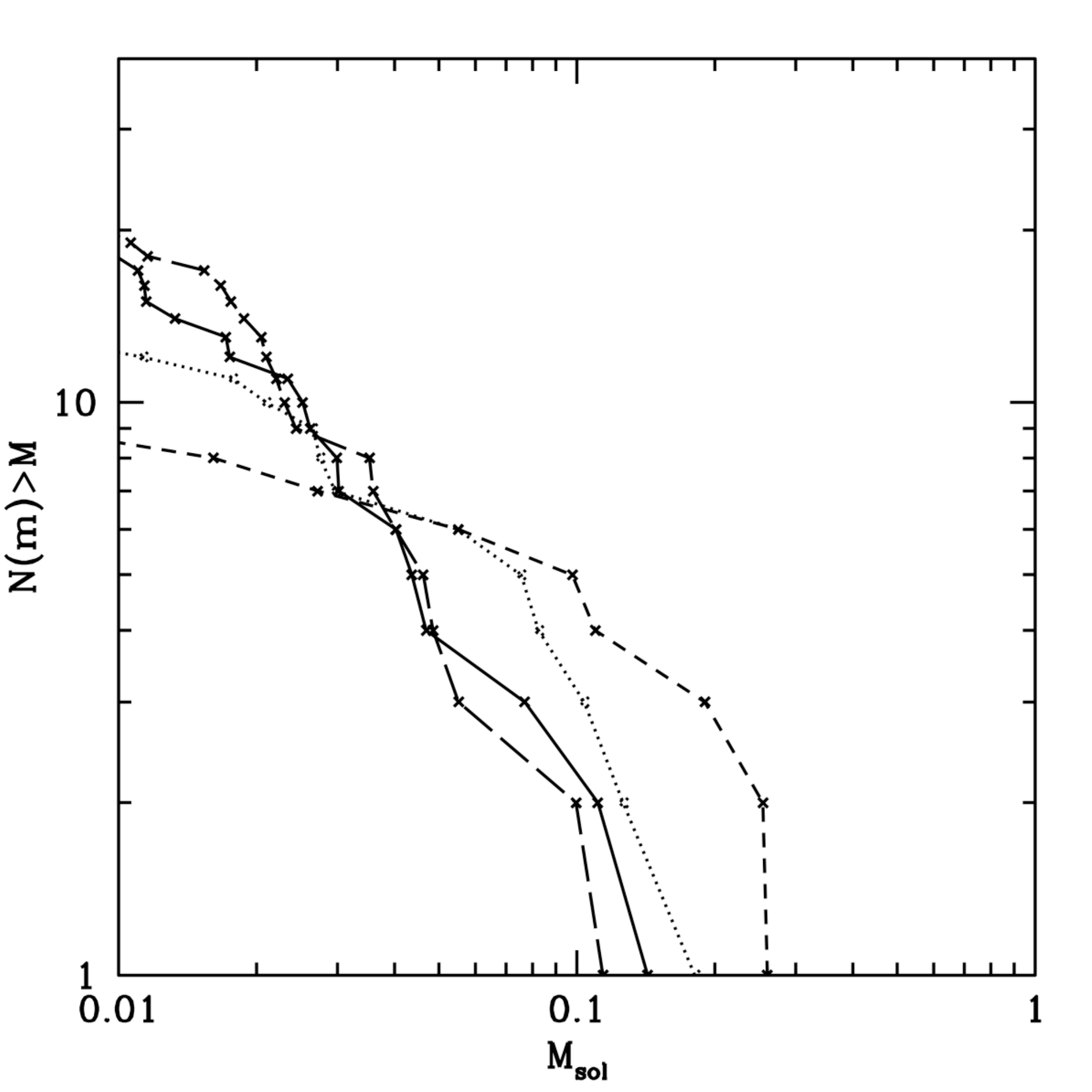} \\
\caption{The variation of the cumulative number clump mass function with resolution. The clump-finds shown are from the \textit{dashed line} $50\times50$ grid with a cell width of $4,125$ AU, \textit{dotted line} $100\times100$ grid with a cell width of $2,063$ AU, the \textit{solid line} $200\times200$ grid with a cell width of $1,031$ AU and the \textit{long dashed line} $400\times400$ grid with a cell width of $516$ AU.}
\label{rescmf}
\end{center}
\end{figure}

Our analysis of the PP clump-finding procedure starts with examining a fundamental difficulty in interpreting the observed clump properties: would the results differ if the observed cloud was further away? We address this problem by changing the resolution of the grid onto which we interpolate our SPH data. Perhaps unsurprisingly, we find that the number of clumps identified in the data varies with the resolution of the grid. When the same physical region of the SPH data is interpolated to a grid with smaller elements, the clump-finding procedure yields more clumps. Figure \ref{Resolution} shows the clumps extracted by the algorithm from 4 data grids of the following sizes, $50\times50$, $100\times100$, $200\times200$ and $400\times400$ grid cells . Table \ref{Restab} shows the number of clumps found in each case and the spatial resolutions in physical units. Note that these resolutions are comparable to current observational surveys of nearby star-forming regions. Figure \ref{rescmf} shows the clump mass functions. 

As the resolution of the clump-finding input increases the slope of the high mass end steepens due to more intermediate-mass objects being found and the biggest objects being broken into smaller units. Additionally the break position between the two regimes decreases as the resolution increases. The slope of the low mass end does not appear to change significantly, however the limited number of clumps in this regime, particularly in the low resolution cases, make it hard to draw definite conclusions.

\begin{table}
	\centering
	\caption{The recovered clumps at different resolutions for clump-find on PP data. The * denotes our fiducial case and the resolution is given in grid cells.}
		\begin{tabular}{l c c c}
   	         \hline
	         \hline
	         Resolution & Spatial Resolution & No. Clumps & Total Mass\\
	         \hline
	         $50\times50$ & $4,125$ AU &$12$ & $26.04$ M$_{\odot}$\\
	         $100\times100$ & $2,063$ AU & $17$ & $18.62$ M$_{\odot}$\\
	         $200\times200$ *& $1,031$ AU & $28$ & $16.35$ M$_{\odot}$\\
	         $400\times400$ & $516$ AU & $31$ & $15.71$ M$_{\odot}$\\
		\hline
		\end{tabular}
	\label{Restab}
\end{table}

The amount of mass found in clumps decreases as the number of clumps and the resolution increases. There are two reasons for this. First, the higher resolution cases allow one to resolve peaks in the structure that are blended at lower resolution. Secondly, we define our clumps down to the lowest contour that is shared with a neighbour. As such, two newly resolved smaller clumps that were previously one larger clump, will have less mass than their larger counterpart. Note that in the standard CLUMPFIND algorithm, the total mass would be the same, since the clumps fill the entire surface/volume of the data grid. However a similar analysis performed with CLUMPFIND on observational data may still yield a similar result to what we find here, since the background is usually subtracted before the clump-finding is applied.

\subsection{Orientation}
\label{subsec:blend}
\begin{figure}
\begin{center}
\includegraphics[width=3in]{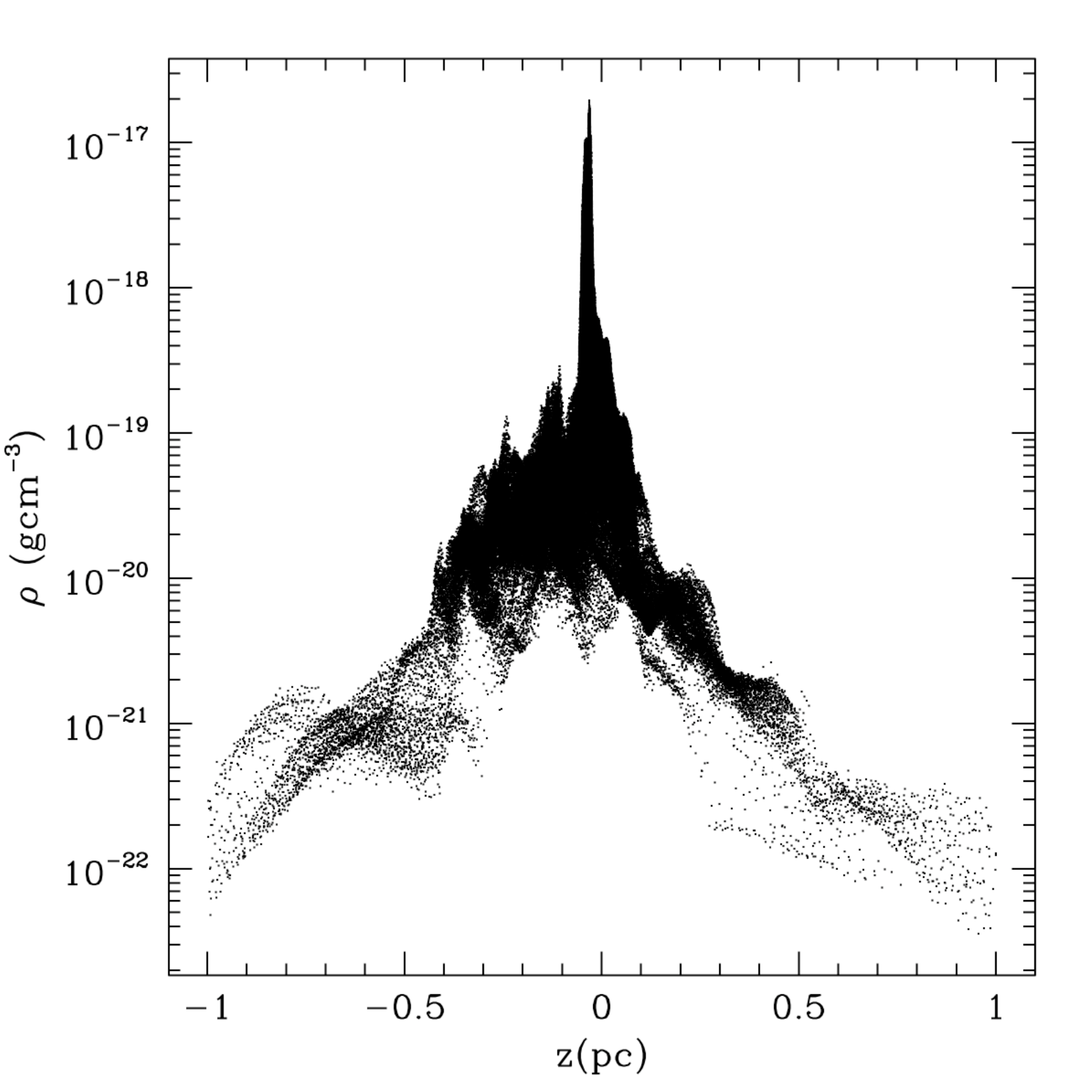} \\
\caption{The intrinsic 3D density profile along the z axes of the material belonging to a typical clump in the standard PP clump-find. About half the material assigned to the clump is truly at high 3D densities but the rest is contamination from material along the line of sight.}
\label{Contamination}
\end{center}
\end{figure}

A further difficulty in interpreting the results of PP clump-finding is that a three dimensional structure has been arbitrarily projected into two dimensions, hence, it is unclear whether a clump is one object, or a superposition of several objects along the line of sight (\citealt{Ballesteros-Paredes02}). Figure \ref{Contamination} shows position along the line of sight and the real 3D density of the material contributing to a typical dense clump in the standard PP clump-find. The material is too widely distributed to be a single core. Moreover, some of this gas lies at densities (3D) which actually sufficiently preclude it belonging to any dense core.

\begin{figure}
\begin{center}
\includegraphics[width=3in]{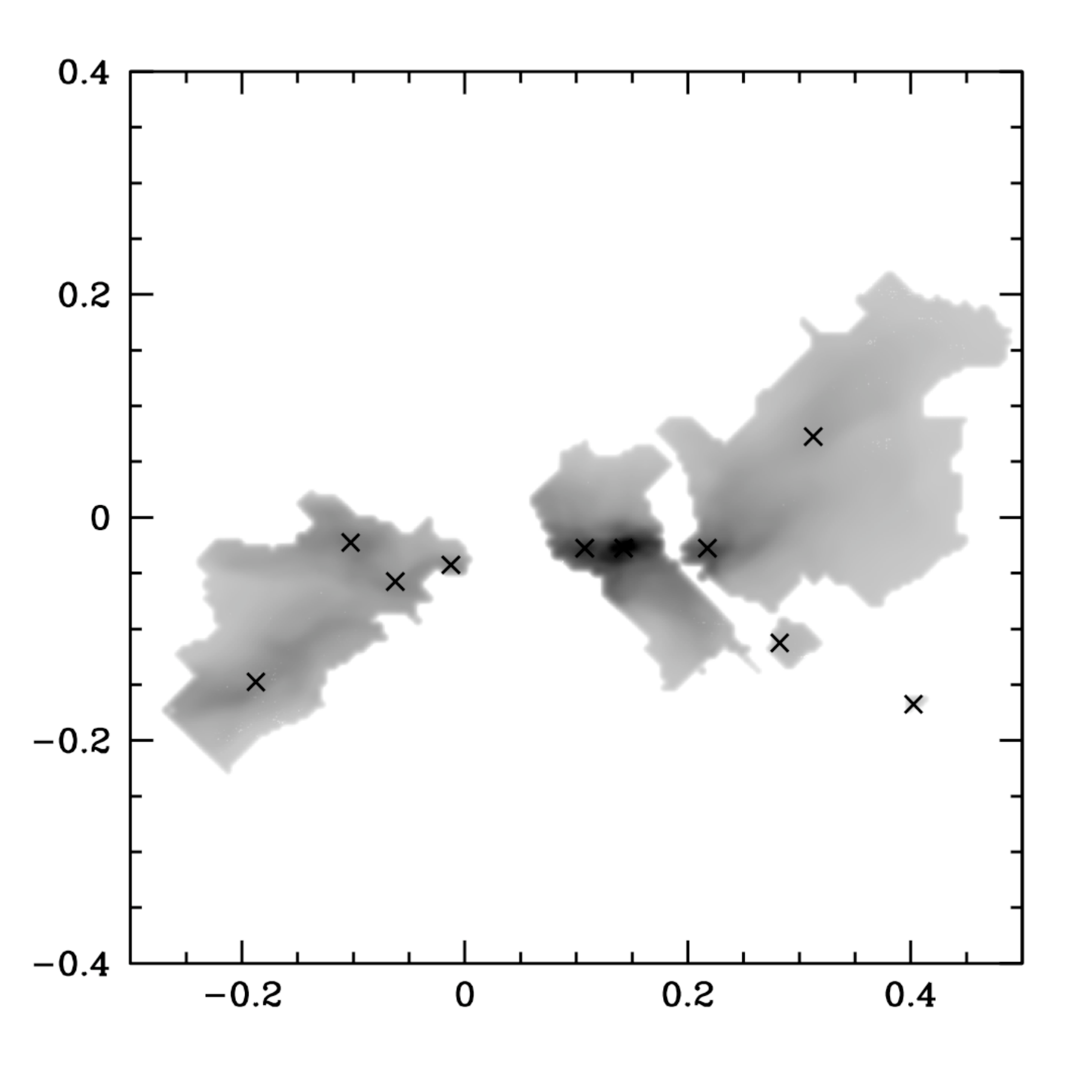}
\caption{A clump-find with identical parameters to our standard PP clump-find carried out on the simulated data projected into the xz plane instead of the xy plane. The scale is shown in parsecs and the grayscale represent column densities in the range $0.02$ gcm$^{-2}$ (grey) to $1$ gcm$^{-2}$ (black) at logarithmic intervals. Crosses show the centre of the clumps}
\label{Orientation}
\end{center}
\end{figure}

Figure \ref{Orientation} shows that line of sight blending causes different clumps to be found if the three dimensional data set is projected along a different axes. Table \ref{Orientab} shows that different orientations result in different clump properties. This is in agreement with previous work by \citet{Ostriker01}.

\begin{table}
	\centering
	\caption{The properties of the clumps found in different orientations. The starred case corresponds to our standard PP clump-find.}
		\begin{tabular}{l c c}
   	         \hline
	         \hline
	         Projection & No. Clumps & Total Mass (M$_{\odot}$)\\
	         \hline
	         xy *& $28$ & $16.35$ \\
	         xz & $10$ & $27.33$ \\
		\hline
		\end{tabular}
	\label{Orientab}
\end{table}

\subsection{Density Cut}

\begin{figure}
\begin{center}
\includegraphics[width=3in]{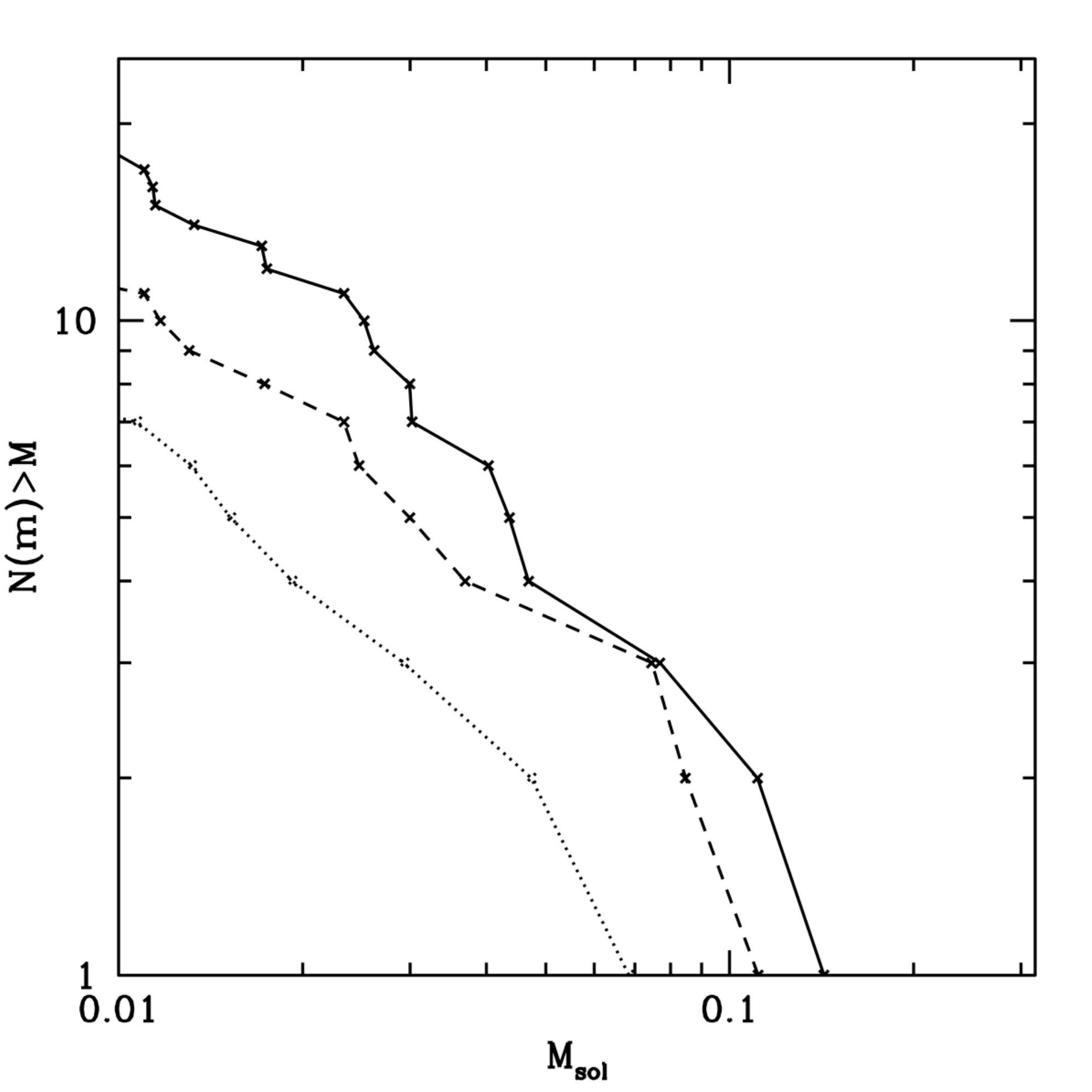}
\caption{The cumulative number clump mass functions found from PP data with three different lower column density limits; \textit{solid line}  $0.01$ gcm$^{-2}$, \textit{dashed line} $0.05$ gcm$^{-2}$ and \textit{dotted line} $0.1$ gcm$^{-2}$. }
\label{drcmf}
\end{center}
\end{figure}

The result of a clump-find is also dependent on the density range of the data-set. Figure \ref{drcmf} shows clump mass functions obtained from a basic PP data-set with three different lower-density sensitivities. As the density is lowered more clumps are discovered and a greater fraction of the structure is probed.

With a restricted density range only the high mass end of the CMF is found. Then when the sensitivity is lowered further the CMF resembles two power laws. 

If the PP data spans a narrower density range the resulting clumps are more distinct. Due to their decreased size, they contain less mass, in other words only the dense centers of the clumps are now included. The second panel of Figure \ref{Compare} shows what the standard PP clump-find becomes if applied to gas with a number density above $10^4$ cm$^{2}$.

\section{Position-Position-Velocity (PPV) clumpfinding}

\label{subsec:den}
\begin{figure*}
\begin{center}
\begin{tabular}{c c}
\includegraphics[width=3in]{respos73c.pdf} &
\includegraphics[width=3in]{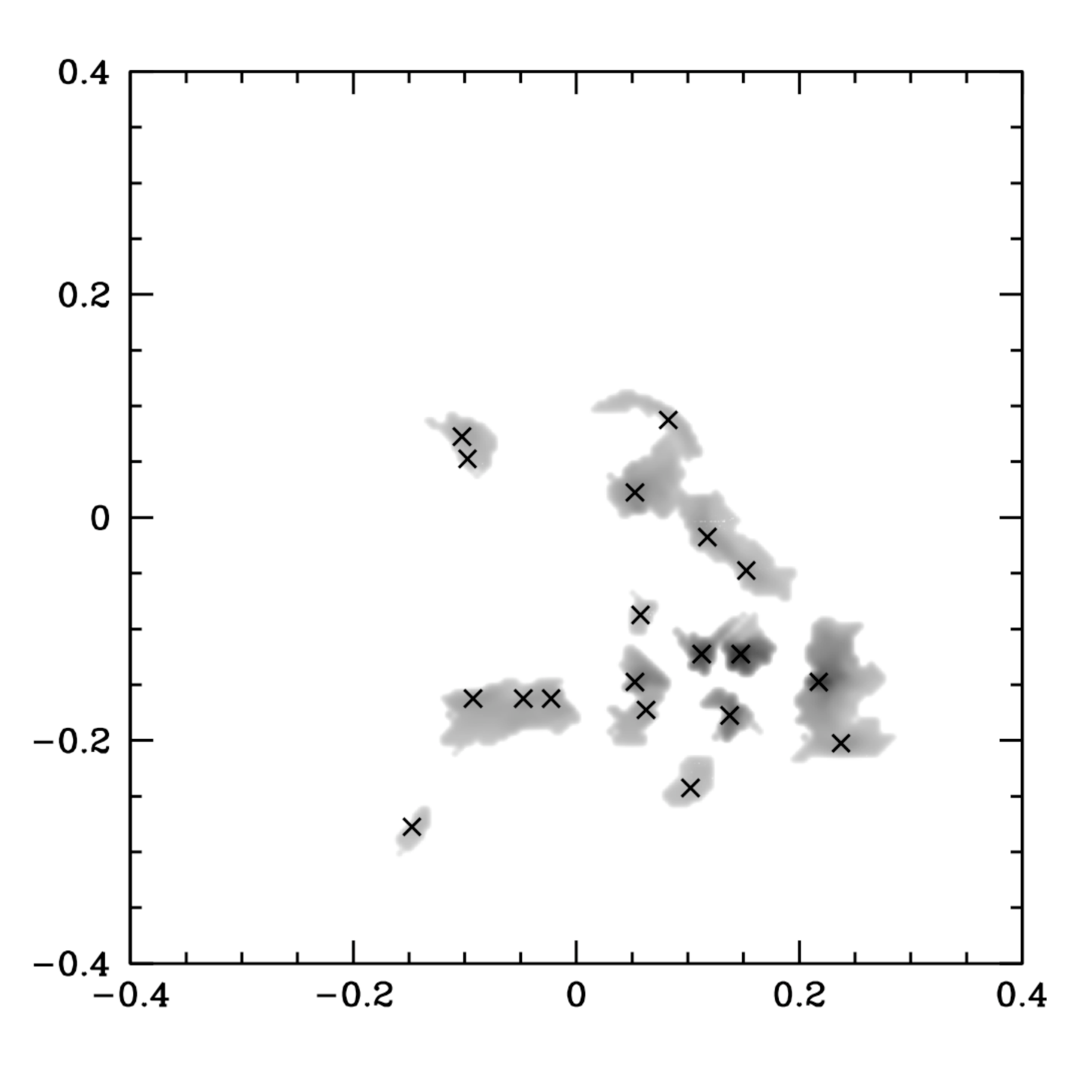} \\
\includegraphics[width=3in]{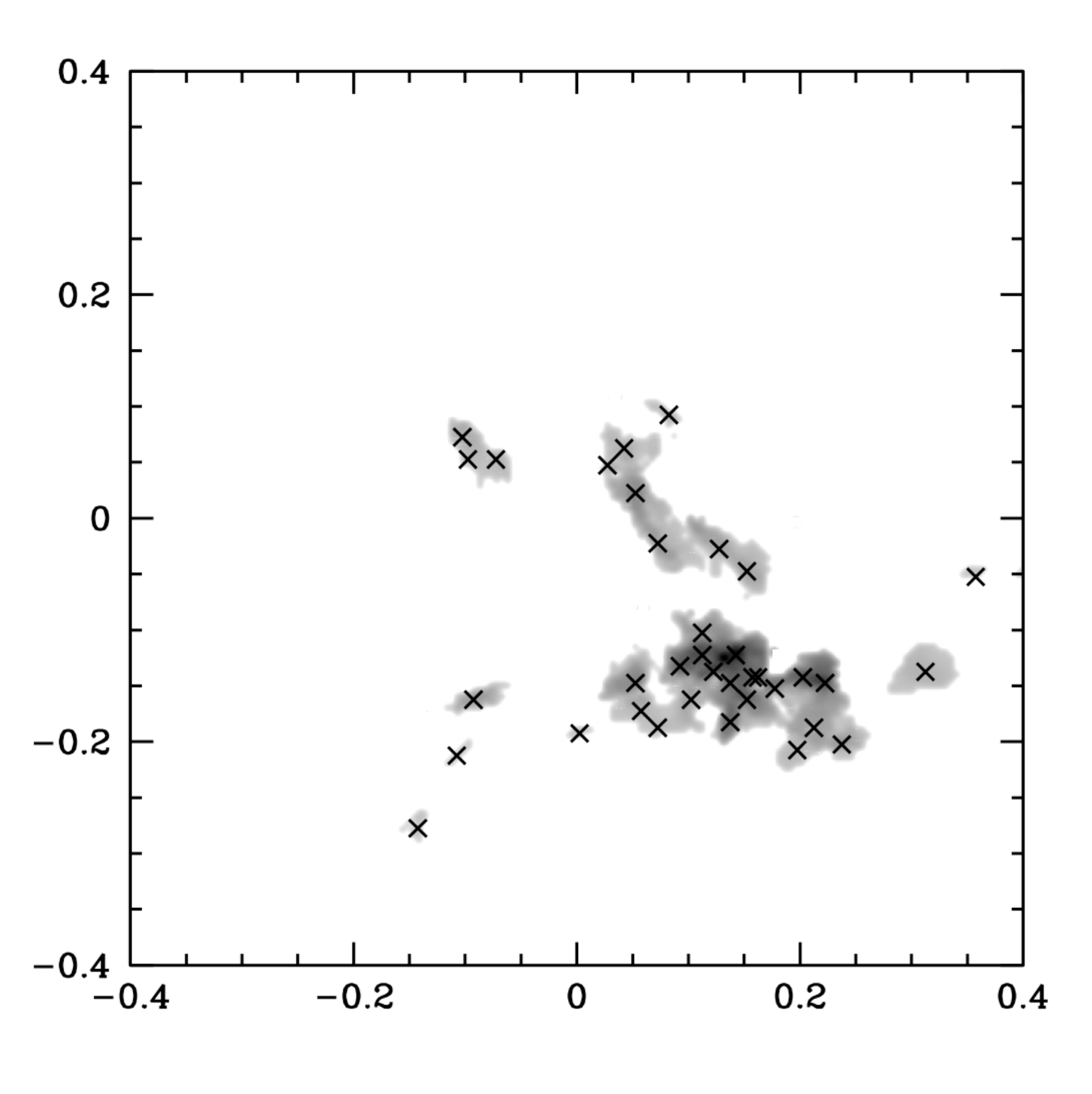} &
\includegraphics[width=3in]{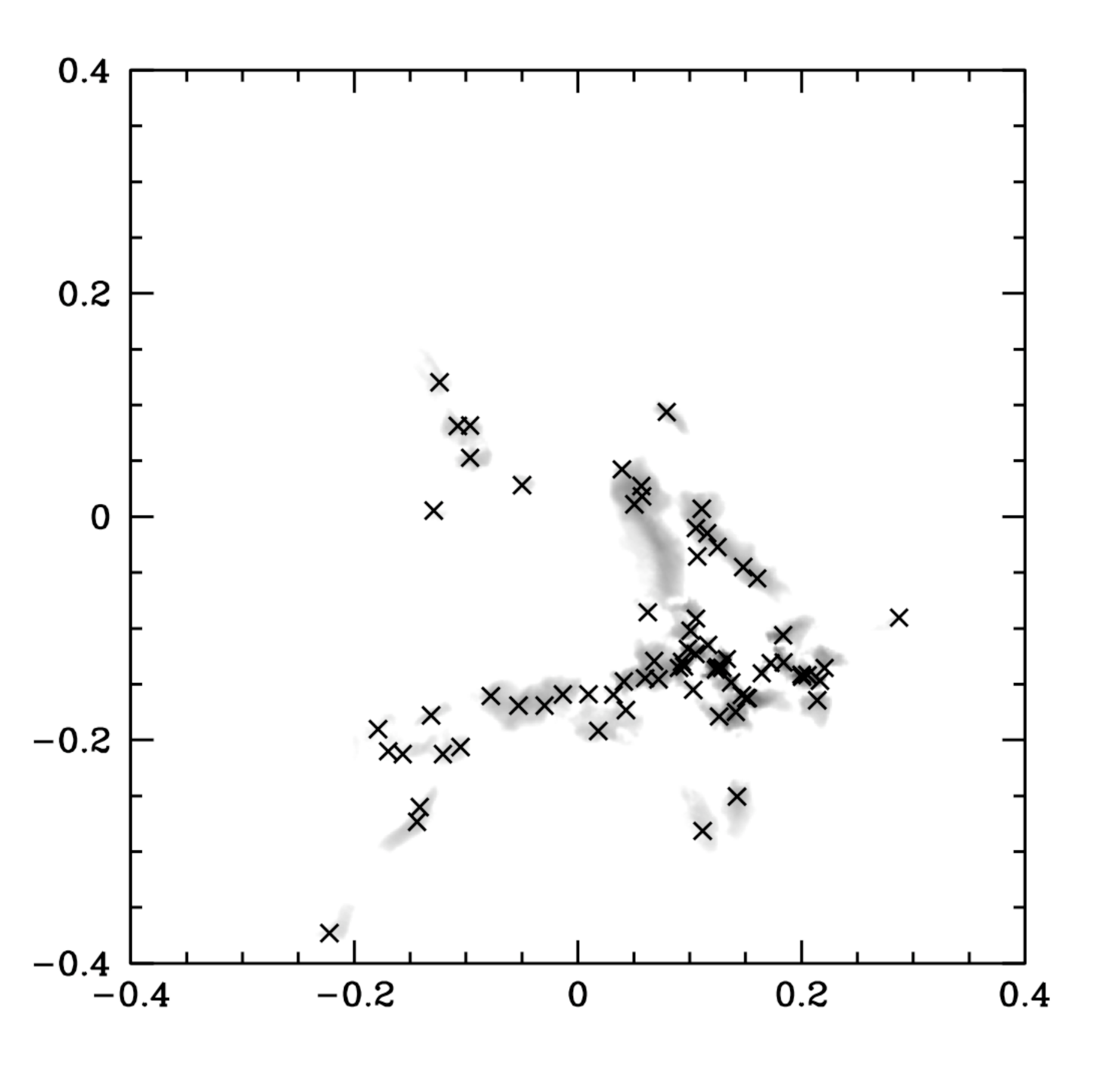}\\
\end{tabular}
\caption{Clumpfinds using various data-sets.\textit {Top left} the standard PP clump-find. \textit{ Top right} applied to a data set above an intrinsic 3D density of  $10^{4}$ cm$^{-3}$. \textit{Bottom left} using the PPV data-set with $16$ velocity bins, corresponding to a velocity resolution of $0.33$ kms$^{-1}$. \textit{Bottom right} the 3D clumpfind on the raw SPH data. The scale is shown in parsecs and the grayscale represent column densities in the range $0.02$ gcm$^{-2}$ (grey) to $1$ gcm$^{-2}$ (black) at logarithmic intervals. Crosses show the centre of the clumps.}
\label{Compare}
\end{center}
\end{figure*} 

The second type of data set we consider in our analysis is position-position column density maps with the radial velocity used as an additional dimension. This data set should suffer less from line of sight blending as it is likely that different clumps will be travelling at different velocities. The width of the velocity channels used determine the velocity resolution of the clumpfind, in our analysis we assign $16$ velocity bins along the z-axes of our data cube which corresponds to a resolution of $0.33$ kms$^{-1}$. Due to the increased number of grid cells a PPV clumpfind uses, it has an intrinsically higher resolution than a PP clumpfind. Panel three of Figure \ref{Compare} shows the PPV clumpfind, and Table \ref{PPVtab} compares its properties to the standard PP case and to an additional PPV case with $40$ velocity bins.

\begin{table}
	\centering
		\begin{tabular}{l c c c}
   	         \hline
	         \hline
	          Velocity Bins & Resolution & No. Clumps & Total Mass\\
	         \hline
	         $0*$  & $-$ & $30$ & $16.37$ M$_{\odot}$ \\
	         $16$ & $0.33$ kms$^{-1}$&$37$ & $8.90$ M$_{\odot}$\\
	         $40$ &  $0.13$ kms$^{-1}$&$33$ & $4.50$ M$_{\odot}$\\
		\hline
		\end{tabular}
	\caption{The properties of the clumps using the PPV data-set compared to the standard PP clump-find. The $16$ velocity bins have a width of $0.33$ kms$^{-1}$ and the $40$ velocity bins a width of $0.13$ kms$^{-1}$}
	\label{PPVtab}
\end{table}

In the PPV case, clump masses are reduced as material previously included is rejected due to its velocity. The positions and extent of the clumps also change relative to the PP clumpfind. Changing the size of the velocity bins also changes the mass and number of clumps found. In the case shown here, when the bin size is decreased and hence the resolution is increased, there is less mass assigned to the clumps. A side effect of this process is that some of the clumps masses decrease to the point that they no longer meet the resolution requirement and are therefore not included in our analysis.

\begin{figure}
\begin{center}
\includegraphics[width=3in]{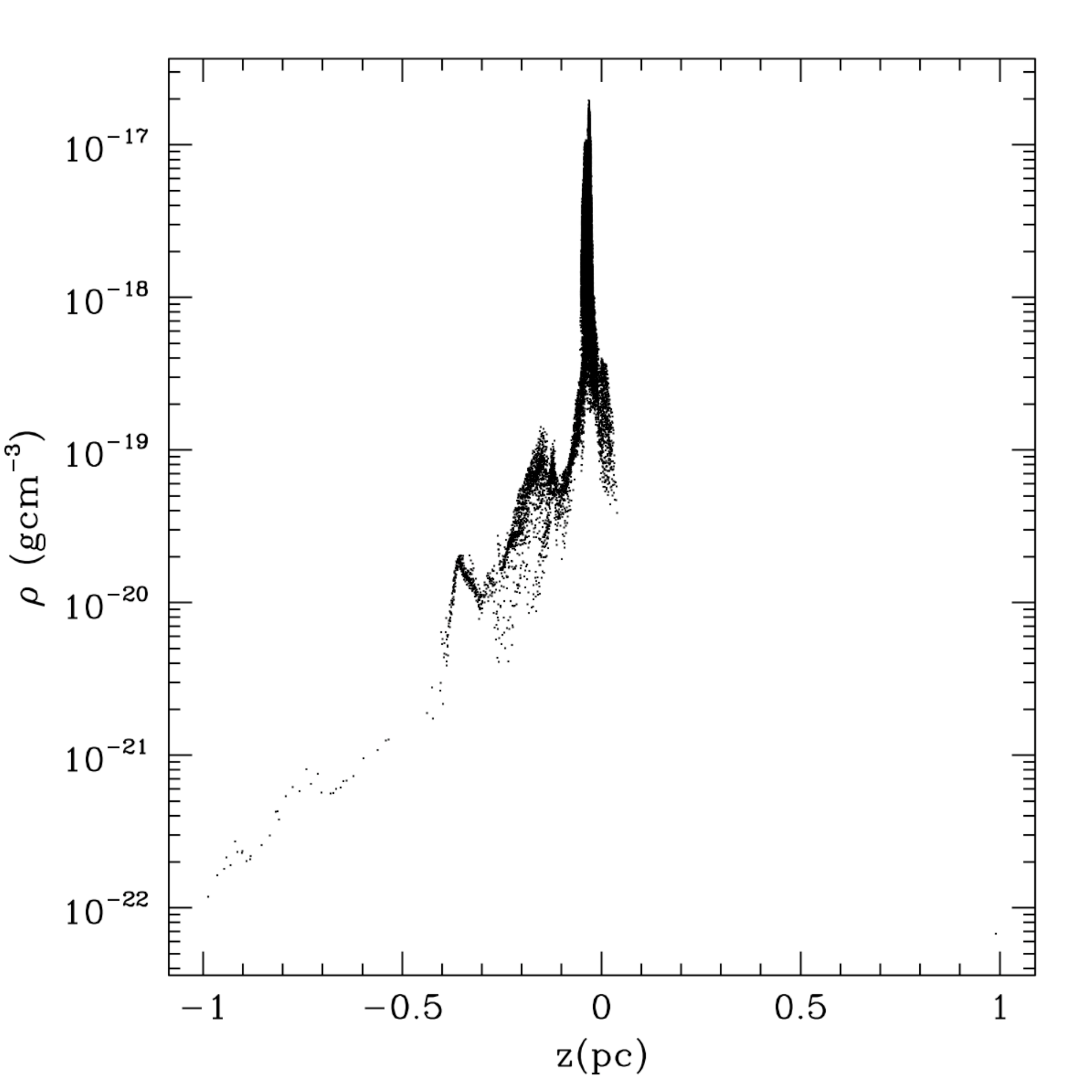}
\caption{A projection of the real 3D density profile of the clump shown in the lower panel of Fig \ref{Contamination} when found using PPV data with 16 velocity bins.}
\label{PPVcontam}
\end{center}
\end{figure}

As regards blending, if we consider the clump found using the PP method shown in Figure \ref{Contamination}, Figure \ref{PPVcontam} shows that more of the mass is now contained in a smaller region. However, the density profile suggests that the clump found still contains spurious material that happens to be moving towards the centre of the region with a similar velocity. Nonetheless, the contribution from material at low density along the line of sight has been reduced. 

It should be noted that the analysis presented here for the PPV method is a `best case scenario' as no molecular tracer can actually probe the entire line of sight from the low density outer layers to the dense core centre. This means that the velocity information is likely to be incomplete. Further, temperature gradients in the core centres will also affect the line profiles. An additional caveat of the PPV method is that if the clump is undergoing collapse or expansion then it could be split into two objects depending on the velocity resolution. However this is still a useful observational technique, particularly when using multiple molecular tracers (e.g. \citealt{Andre07}).

\section{Three-dimensional (PPP/SPH) clumpfinding}

The final data set we consider is that of our full simulated three-dimensional structure. While this is impossible from observations, it is still a useful comparison as the true structure of star forming regions is, of course, three-dimensional. The same basic method is used to find the clumps as before. This is possible as in SPH forces are calculated using a hierarchical tree, consequently a neighbours list exists which allows the densities of neighbouring particles to be compared in a similar manner to grid cells.

However, unlike the PP and PPV clump-finds the resolution is no longer simply the size a grid. In SPH simulations the mass resolution varies spatially, high density regions are better resolved than low ones, therefore our three dimensional clumpfind also shares this property. This effectively results in our SPH run having the highest resolution of all the clumpfinding methods.

The bottom right panel of Fig. \ref {Compare} shows the result of the clump-find on the 3D SPH data. The number of clumps has increased and the total mass decreased relative to our previous results. There are two reasons for this. First, due to our improved resolution the number of clumps increases as shown in Section \ref{subsec:res}. Secondly the use of 3D data eliminates the blending along the line of sight previously encountered in Section \ref{subsec:blend} so more clumps are found with smaller masses. Effectively we are now detecting a finer level of structure than in the previous data-sets, namely the very dense core centers.

\section{Discussion of Global Properties}

\subsection{ The Clumps}

Table \ref{SPHtab} compares the properties of the clump-finds. Although the general outline of the dense region where the clumps are identified remains constant, the number, masses and assigned boundaries of the clumps varies with extraction method. Generally the number of clumps increases and the total mass found in clumps decreases with the precision of the method.

Applying a density cut to a data-set produces more distinct clumps, but is of course completely arbitrary. The PPV clump distribution is a better match to the high-resolution 3D clumpfind than the PP clumps. The peak positions identified as the heads of clumps from the PPV data correlates well with the highest density structures seen in the sph case. The overall mass distribution and the clump properties, however, differ. Each data-set appears to highlight a different scale of the structure in our simulation. 

\begin{table}
	\centering
		\begin{tabular}{l c c}
   	         \hline
	         \hline
	          Method & No. Clumps & Total Mass ($M_{\odot}$)\\
	         \hline
	         PP  & $28$ & $16.35$ \\
	         PP high density & $19$ & $10.51$ \\
	         PPV $16$ bins & $37$ & $8.90$ \\
	         sph & $72$ & $4.50$\\
		\hline
		\end{tabular}
	\caption{A comparison of the clump-find results.}
	\label{SPHtab}
\end{table}

Defining the clump boundaries is further complicated by arbitrary decisions that have to be made in the clumpfinding process. Is there a minimum density required for a clump? How many density contour levels need to be set? Should some minimum contrast to the background be required? All these issues add to the difficulties presented here in comparing clumps found from different data-sets. These challenges should be kept in mind when comparing the properties of cores recorded in the literature, particularly when contrasting masses and sizes.

\subsection{The Clump Mass Function}

\begin{figure*}
\begin{center}
\begin{tabular}{c c}
\includegraphics[width=2.2in]{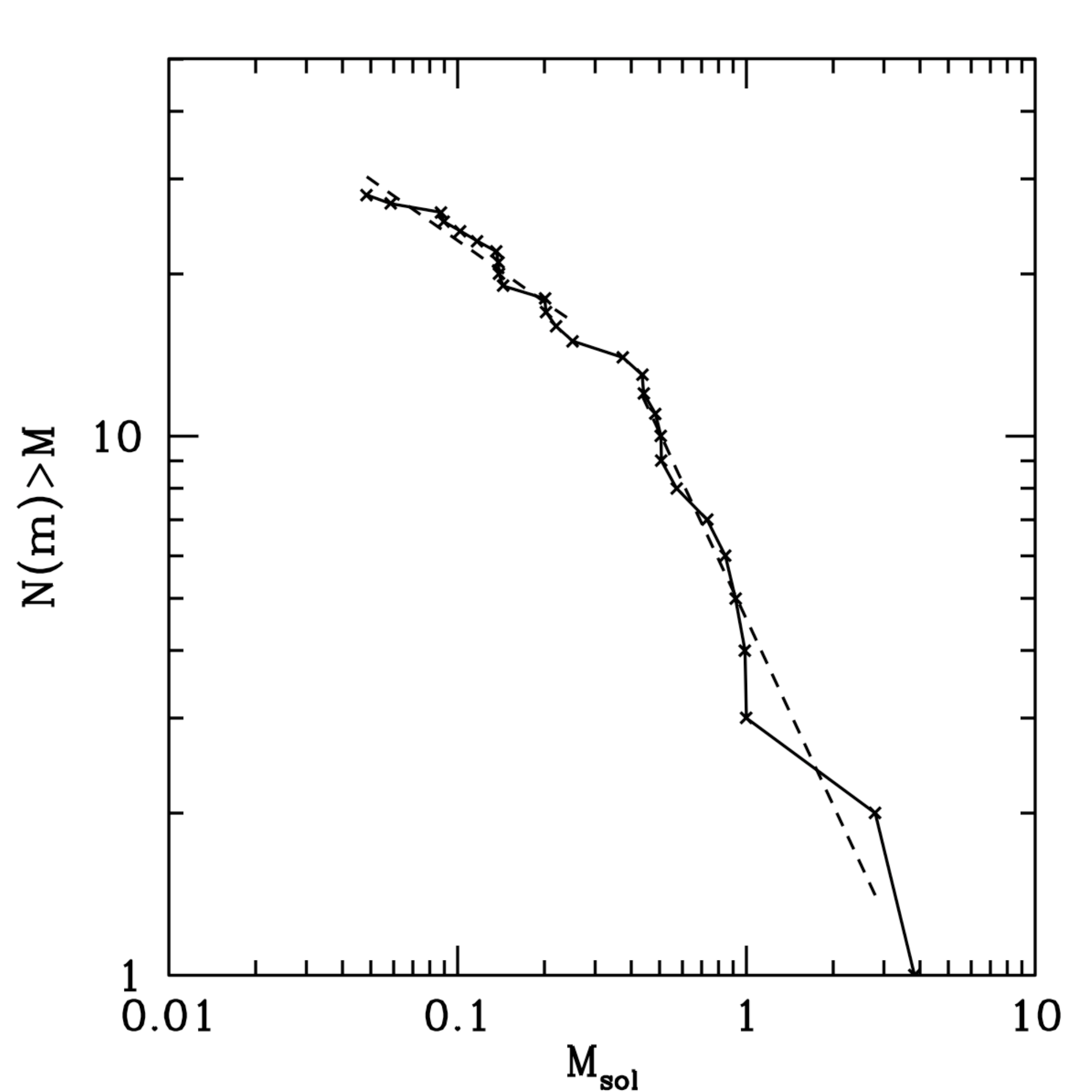} &
\includegraphics[width=2.2in]{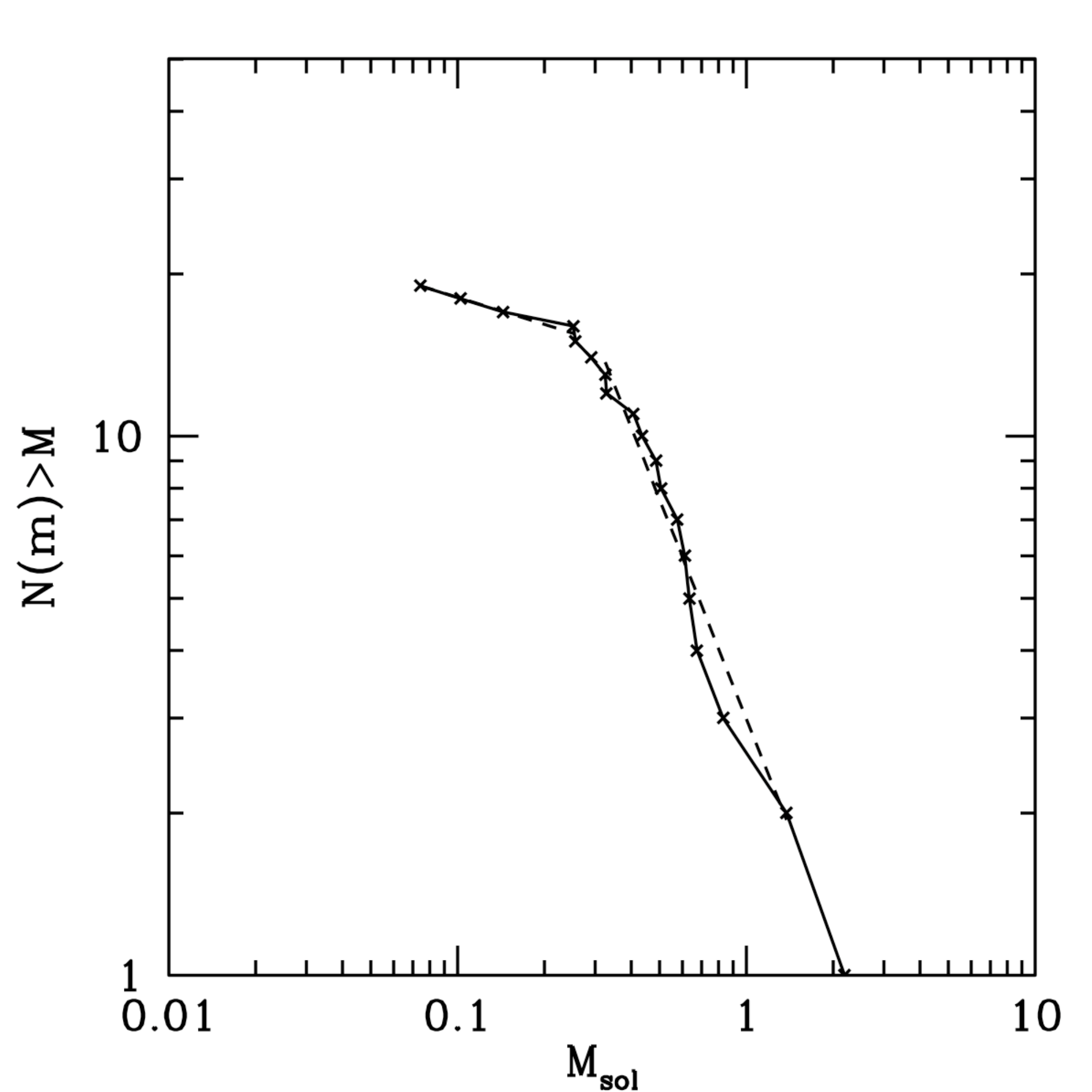} \\
\includegraphics[width=2.2in]{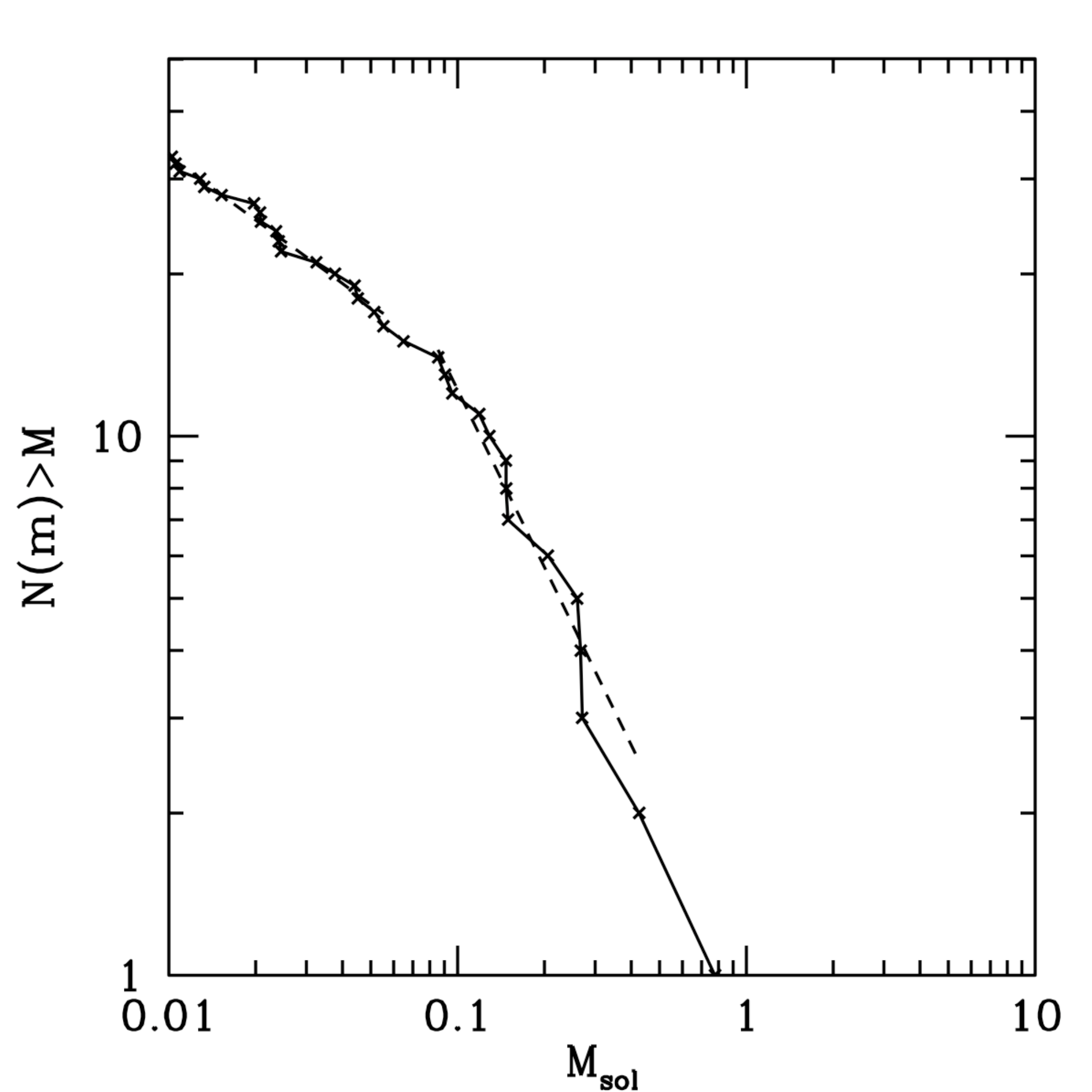} &
\includegraphics[width=2.2in]{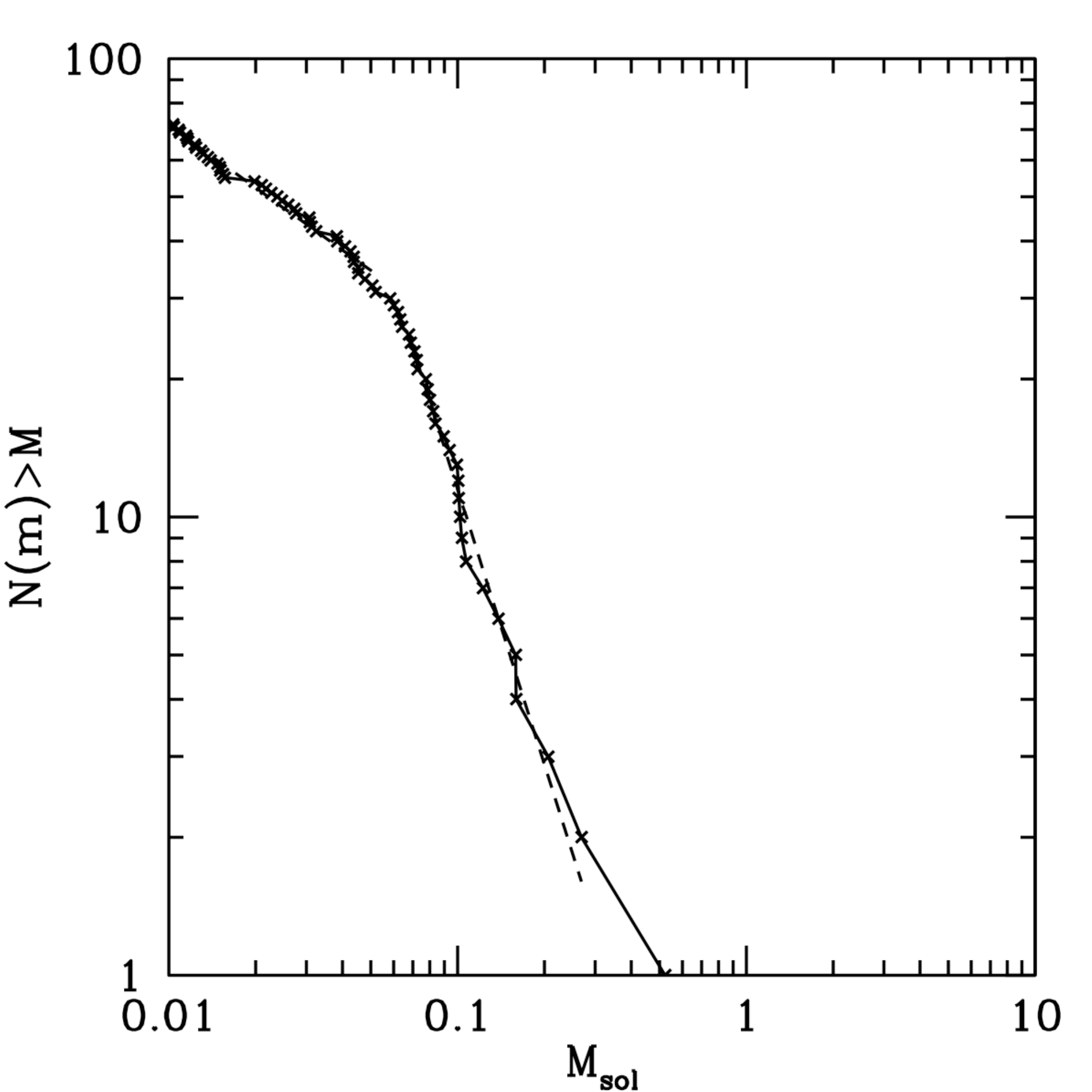} \\
\end{tabular}
\caption{The cumulative number clump mass functions from the previous clumpfinds. \textit{Top left}  our standard PP clump-find, \textit{top right} high density PP data, \textit{bottom left}  the PPV clump-find  and \textit{bottom right} the PPP (SPH) method. The dashed lines are power law fits whose gradients are shown in Table \ref{fits}.}
\label{cmfs}
\end{center}
\end{figure*}

Despite the  differences in the clump properties, the clump mass functions show marked similarities. Figure \ref{cmfs} shows the cumulative clump mass functions derived from the PP, PP with density cut, PPV and PPP (SPH) clump-finds and Table \ref{fits} shows the gradients of the power laws which best fit them.

\begin{table}
	\centering
		\begin{tabular}{l c c}
   	         \hline
	         \hline
	          Method & Low Mass Gradient & High Mass Gradient\\
	         \hline
	         PP  & $-0.37$ & $-1.15$ \\
	         PP high density & $-0.16$ & $-1.35$ \\
	         PPV $16$ bins & $-0.39$ & $-1.10$ \\
	         sph & $-0.45$ & $-1.99$\\
		\hline
		\end{tabular}
	\caption{The best fit power law gradients for the cumulative CMF's shown in Figure \ref{cmfs}.}
	\label{fits}
\end{table}

As the clump properties differ each of the CMF's has a different magnitude, size range and break point but the overall shape is always broadly consistent with the stellar initial mass function \citep{Kroupa02}. When plotted logarithmically this corresponds to a Saltpeter slope of $\Gamma=-1.3$ with a break to second power law of $\Gamma=-0.3$ at lower masses. 

At the low mass end, with the exception of the PP high density case, all the gradients are within $0.15$ of the Kroupa value and within $0.08$ of each other. The PP case where the density range was restricted to high values resulted in incomplete sampling of the low-mass end, which leads to a flattening of the low-mass gradient. At the high mass end the two-dimensional methods have gradients within $0.25$ of each other, poisson noise makes these fits more uncertain but the agreement is still good. The best fit to the high mass end of the SPH cumulative clumps mass function, however, is steeper than the two dimensional cases. It is unclear whether this is a genuine consequence of the 3D data set or due to the increasing resolution of the SPH method with density preferentially identifying smaller denser structures.

Further evidence that the CMF may have a universal shape can be seen in our earlier discussions of resolution (Figure \ref{rescmf}) and density range (Figure \ref{drcmf}). Once again in these cases we see that although the properties of the clumps are changing, the profile of the CMF remains broadly consistent with the stellar IMF.

These results imply that the shape of the CMF results naturally from breaking up the underlying structure of a molecular cloud in a consistent manner. Therefore, the emergence of a rational CMF from a data set is no guarantee that the identified clumps are unique objects. The structure could have been divided up using a differing, perfectly justifiable, method which results in different clump positions and sizes but yields a similar CMF profile. This is reminiscent of the observations discussed in the introduction, all have similar CMFs but the characteristic masses of clumps differ. Indeed, due to difficulties like those discussed here, \citet{Rosolowsky08} has proposed that dendrograms might ultimately prove a better way to identify structure than the traditional clumps. Still, many of the issues raised here are likely to be relevant using dendrograms or any other clump finding algorithm.

\subsection{Universality of the CMF}

Despite variations in clump properties, the slope of the clump mass function is similar in all the cases investigated here, with the possible exception of the high mass end in the SPH clump mass function. The CMF retains this shape when extracted using different clump-finding methods and parameters.

As many studies have shown (e.g. \citealt{Vazquez-Semadeni95}; \citealt{Ballesteros-Paredes99}; \citealt{Klessen01}), turbulence generates the characteristic multi-scalar structure of molecular clouds. \citet{Elmegreen02} goes so far as to describe the structure of the interstellar medium as a continuous fractal distribution.  Observationally, \citet{Padoan03} show that the structure of the Taurus and Perseus molecular clouds is hierarchical at scales of $0.3-3$ pc. Given the universality of turbulence \citep{Heyer04}, it is perhaps unsurprising that the observed CMFs exhibit similarities between different star forming regions, and there exists in the literature a number of models for describing how these clumps can form  \citep*{Fleck82, Elmegreen93, Padoan95, Padoan97,Myers00, Klessen01, Padoan02}. The similarities between the CMF and the IMF then naturally leads to the suggestion that the two are related  \citep*{Motte98, Lada08, Myers08}.

However, our findings suggest that the mapping between the CMF and IMF is not straightforward. Analysing fractal-like structure with any clump-finding algorithm will generate a  clump mass function. The details of the clump properties, especially the mass, will depend on the type of algorithm and the parameters chosen. Further, the clumps are not uniquely defined. This has also been found observationally. \citet{Johnstone00} and \citet{Motte98} both surveyed the $\rho$ Ophiuchi and derived clump mass functions which were in agreement with each other. However, as \citet{Johnstone00} comment, the masses, numbers and positions of the clumps found do not correspond between the surveys. Due to the variance of clump properties with identification method, it is unclear how best to link star-less cores to protostars in a quantitative manner. 

\section{Conclusions}

We have shown that when a CLUMPFIND style algorithm is applied to different types of data-sets from the same underlying simulation, clump characteristics vary. This is due to a number of factors. 
Details of the observational maps or the simulation obviously play their part, particularly parameters  such as resolution, orientation and density range. The more levels of structure resolved within a data-set the greater the number of clumps that can be found. Blending of structure along the line of sight decreases the number of clumps found in 2D clumpfinds and increases clumps masses. A significant amount of material which is in reality not connected to a high density core can be artificially included. Applying a high density cut to PP data yields more distinct objects but can bias the resulting mass function.

Using velocity as a third dimension improves accuracy and provides a closer match to the true 3D structure than just using PP data. The clump boundaries, however, still differ and less mass is assigned to clumps.

Qualitatively the profile of the clump mass function always resembles the Stellar IMF, regardless of the data-set used. However, the quantitative values of the CMF are dependent on the extraction method, and therefore causal comparisons to the IMF should be made with caution. 

\section{Acknowledgements}

We would like to thank Jane Greaves, Thomas Maschberger, Alyssa Goodman, Katharine Johnstone and Thomas Robitaille for useful discussions. We would also like to acknowledge the constructive comments made by an anonymous referee.  P.C.C. acknowledges support by the Deutsche 
Forschungsgemeinschaft (DFG) under grant KL 1358/5 and via the 
Sonderforschungsbereich (SFB) SFB 439, Galaxien im fr\"uhen Universum. Lastly, we would like to acknowledge the Scottish Universities Physics Alliance (SUPA) for providing the computing facilities required for this work. 

\bibliography{Bibliography}

\begin{thebibliography}{}

\bibitem[\protect\citeauthoryear{{Alves}, {Lombardi} \& {Lada}}{{Alves}
  et~al.}{2007}]{Alves07}
{Alves} J.,  {Lombardi} M.,    {Lada} C.~J.,  2007, \aap, 462, L17

\bibitem[\protect\citeauthoryear{{Andr{\'e}}, {Belloche}, {Motte} \&
  {Peretto}}{{Andr{\'e}} et~al.}{2007}]{Andre07}
{Andr{\'e}} P.,  {Belloche} A.,  {Motte} F.,    {Peretto} N.,  2007, \aap, 472,
  519

\bibitem[\protect\citeauthoryear{{Ballesteros-Paredes} \& {Mac
  Low}}{{Ballesteros-Paredes} \& {Mac Low}}{2002}]{Ballesteros-Paredes02}
{Ballesteros-Paredes} J.,  {Mac Low} M.-M.,  2002, \apj, 570, 734

\bibitem[\protect\citeauthoryear{{Ballesteros-Paredes}, {V{\'a}zquez-Semadeni}
  \& {Scalo}}{{Ballesteros-Paredes} et~al.}{1999}]{Ballesteros-Paredes99}
{Ballesteros-Paredes} J.,  {V{\'a}zquez-Semadeni} E.,    {Scalo} J.,  1999,
  \apj, 515, 286

\bibitem[\protect\citeauthoryear{{Bate}, {Bonnell} \& {Price}}{{Bate}
  et~al.}{1995}]{Bate95}
{Bate} M.~R.,  {Bonnell} I.~A.,    {Price} N.~M.,  1995, \mnras, 277, 362

\bibitem[\protect\citeauthoryear{{Bate} \& {Burkert}}{{Bate} \&
  {Burkert}}{1997}]{Bate97}
{Bate} M.~R.,  {Burkert} A.,  1997, \mnras, 288, 1060

\bibitem[\protect\citeauthoryear{{Benz}}{{Benz}}{1990}]{Benz90}
{Benz} W.,  1990, in {Buchler} J.~R.,  ed., Numerical Modelling of Nonlinear
  Stellar Pulsations Problems and Prospects {Smooth Particle Hydrodynamics - a
  Review}.
pp 269--+

\bibitem[\protect\citeauthoryear{{Bergin} \& {Tafalla}}{{Bergin} \&
  {Tafalla}}{2007}]{Bergin07}
{Bergin} E.~A.,  {Tafalla} M.,  2007, \araa, 45, 339

\bibitem[\protect\citeauthoryear{{Brunt} \& {Mac Low}}{{Brunt} \& {Mac
  Low}}{2004}]{Brunt04}
{Brunt} C.~M.,  {Mac Low} M.-M.,  2004, \apj, 604, 196

\bibitem[\protect\citeauthoryear{{Di Francesco}, {Evans} II, {Caselli},
  {Myers}, {Shirley}, {Aikawa} \& {Tafalla}}{{Di Francesco}
  et~al.}{2007}]{DiFrancesco07}
{Di Francesco} J.,  {Evans} II N.~J.,  {Caselli} P.,  {Myers} P.~C.,  {Shirley}
  Y.,  {Aikawa} Y.,    {Tafalla} M.,  2007, in {Reipurth} B.,  {Jewitt} D.,
  {Keil} K.,  eds, Protostars and Planets V {An Observational Perspective of
  Low-Mass Dense Cores I: Internal Physical and Chemical Properties}.
pp 17--32

\bibitem[\protect\citeauthoryear{{Dubinski}, {Narayan} \&
  {Phillips}}{{Dubinski} et~al.}{1995}]{Dubinski95}
{Dubinski} J.,  {Narayan} R.,    {Phillips} T.~G.,  1995, \apj, 448, 226

\bibitem[\protect\citeauthoryear{{Elmegreen}}{{Elmegreen}}{1993}]{Elmegreen93}
{Elmegreen} B.~G.,  1993, \apjl, 419, L29+

\bibitem[\protect\citeauthoryear{{Elmegreen}}{{Elmegreen}}{2002}]{Elmegreen02}
{Elmegreen} B.~G.,  2002, \apj, 564, 773

\bibitem[\protect\citeauthoryear{{Fleck} Jr.}{{Fleck}}{1982}]{Fleck82}
{Fleck} Jr. R.~C.,  1982, \mnras, 201, 551

\bibitem[\protect\citeauthoryear{{Gingold} \& {Monaghan}}{{Gingold} \&
  {Monaghan}}{1983}]{Gingold83}
{Gingold} R.~A.,  {Monaghan} J.~J.,  1983, \mnras, 204, 715

\bibitem[\protect\citeauthoryear{{Goodman}, {Barranco}, {Wilner} \&
  {Heyer}}{{Goodman} et~al.}{1998}]{Goodman98}
{Goodman} A.~A.,  {Barranco} J.~A.,  {Wilner} D.~J.,    {Heyer} M.~H.,  1998,
  \apj, 504, 223

\bibitem[\protect\citeauthoryear{{Heyer} \& {Brunt}}{{Heyer} \&
  {Brunt}}{2004}]{Heyer04}
{Heyer} M.~H.,  {Brunt} C.~M.,  2004, \apjl, 615, L45

\bibitem[\protect\citeauthoryear{{Johnstone}, {Matthews} \&
  {Mitchell}}{{Johnstone} et~al.}{2006}]{Johnstone06}
{Johnstone} D.,  {Matthews} H.,    {Mitchell} G.~F.,  2006, \apj, 639, 259

\bibitem[\protect\citeauthoryear{{Johnstone}, {Wilson}, {Moriarty-Schieven},
  {Joncas}, {Smith}, {Gregersen} \& {Fich}}{{Johnstone}
  et~al.}{2000}]{Johnstone00}
{Johnstone} D.,  {Wilson} C.~D.,  {Moriarty-Schieven} G.,  {Joncas} G.,
  {Smith} G.,  {Gregersen} E.,    {Fich} M.,  2000, \apj, 545, 327

\bibitem[\protect\citeauthoryear{{Klessen}}{{Klessen}}{2001}]{Klessen01}
{Klessen} R.~S.,  2001, \apj, 556, 837

\bibitem[\protect\citeauthoryear{{Klessen} \& {Burkert}}{{Klessen} \&
  {Burkert}}{2000}]{Klessen00}
{Klessen} R.~S.,  {Burkert} A.,  2000, \apjs, 128, 287

\bibitem[\protect\citeauthoryear{{Kroupa}}{{Kroupa}}{2002}]{Kroupa02}
{Kroupa} P.,  2002, Science, 295, 82

\bibitem[\protect\citeauthoryear{{Lada}, {Muench}, {Rathborne}, {Alves} \&
  {Lombardi}}{{Lada} et~al.}{2008}]{Lada08}
{Lada} C.~J.,  {Muench} A.~A.,  {Rathborne} J.,  {Alves} J.~F.,    {Lombardi}
  M.,  2008, \apj, 672, 410

\bibitem[\protect\citeauthoryear{{Larson}}{{Larson}}{1981}]{Larson81}
{Larson} R.~B.,  1981, \mnras, 194, 809

\bibitem[\protect\citeauthoryear{{Mac Low}, {Klessen}, {Burkert} \&
  {Smith}}{{Mac Low} et~al.}{1998}]{MacLow98}
{Mac Low} M.-M.,  {Klessen} R.~S.,  {Burkert} A.,    {Smith} M.~D.,  1998,
  Physical Review Letters, 80, 2754

\bibitem[\protect\citeauthoryear{{Monaghan}}{{Monaghan}}{1992}]{Monaghan92}
{Monaghan} J.~J.,  1992, \araa, 30, 543

\bibitem[\protect\citeauthoryear{{Motte}, {Andre} \& {Neri}}{{Motte}
  et~al.}{1998}]{Motte98}
{Motte} F.,  {Andre} P.,    {Neri} R.,  1998, \aap, 336, 150

\bibitem[\protect\citeauthoryear{{Myers}}{{Myers}}{2000}]{Myers00}
{Myers} P.~C.,  2000, \apjl, 530, L119

\bibitem[\protect\citeauthoryear{{Myers}}{{Myers}}{2008}]{Myers08}
{Myers} P.~C.,  2008, ArXiv e-prints, 807

\bibitem[\protect\citeauthoryear{{Myers} \& {Gammie}}{{Myers} \&
  {Gammie}}{1999}]{Myers99}
{Myers} P.~C.,  {Gammie} C.~F.,  1999, \apjl, 522, L141

\bibitem[\protect\citeauthoryear{{Nutter} \& {Ward-Thompson}}{{Nutter} \&
  {Ward-Thompson}}{2007}]{Nutter07}
{Nutter} D.,  {Ward-Thompson} D.,  2007, \mnras, 374, 1413

\bibitem[\protect\citeauthoryear{{Ostriker}, {Stone} \& {Gammie}}{{Ostriker}
  et~al.}{2001}]{Ostriker01}
{Ostriker} E.~C.,  {Stone} J.~M.,    {Gammie} C.~F.,  2001, \apj, 546, 980

\bibitem[\protect\citeauthoryear{{Padoan}}{{Padoan}}{1995}]{Padoan95}
{Padoan} P.,  1995, \mnras, 277, 377

\bibitem[\protect\citeauthoryear{{Padoan}, {Boldyrev}, {Langer} \&
  {Nordlund}}{{Padoan} et~al.}{2003}]{Padoan03}
{Padoan} P.,  {Boldyrev} S.,  {Langer} W.,    {Nordlund} {\AA}.,  2003, \apj,
  583, 308

\bibitem[\protect\citeauthoryear{{Padoan} \& {Nordlund}}{{Padoan} \&
  {Nordlund}}{2002}]{Padoan02}
{Padoan} P.,  {Nordlund} {\AA}.,  2002, \apj, 576, 870

\bibitem[\protect\citeauthoryear{{Padoan}, {Nordlund} \& {Jones}}{{Padoan}
  et~al.}{1997}]{Padoan97}
{Padoan} P.,  {Nordlund} A.,    {Jones} B.~J.~T.,  1997, \mnras, 288, 145

\bibitem[\protect\citeauthoryear{{Padoan}, {Nordlund}, {Kritsuk}, {Norman} \&
  {Li}}{{Padoan} et~al.}{2007}]{Padoan07}
{Padoan} P.,  {Nordlund} {\AA}.,  {Kritsuk} A.~G.,  {Norman} M.~L.,    {Li}
  P.~S.,  2007, \apj, 661, 972

\bibitem[\protect\citeauthoryear{{Pineda}, {Caselli} \& {Goodman}}{{Pineda}
  et~al.}{2008}]{Pineda08}
{Pineda} J.~E.,  {Caselli} P.,    {Goodman} A.~A.,  2008, ArXiv e-prints, 802

\bibitem[\protect\citeauthoryear{{Rosolowsky}, {Pineda}, {Kauffmann} \&
  {Goodman}}{{Rosolowsky} et~al.}{2008}]{Rosolowsky08}
{Rosolowsky} E.~W.,  {Pineda} J.~E.,  {Kauffmann} J.,    {Goodman} A.~A.,
  2008, ArXiv e-prints, 802

\bibitem[\protect\citeauthoryear{{Schnee}, {Bethell} \& {Goodman}}{{Schnee}
  et~al.}{2006}]{Schnee06}
{Schnee} S.,  {Bethell} T.,    {Goodman} A.,  2006, \apjl, 640, L47

\bibitem[\protect\citeauthoryear{{Swift} \& {Williams}}{{Swift} \&
  {Williams}}{2008}]{Swift08}
{Swift} J.~J.,  {Williams} J.~P.,  2008, \apj, 679, 552

\bibitem[\protect\citeauthoryear{{Testi} \& {Sargent}}{{Testi} \&
  {Sargent}}{1998}]{Testi98}
{Testi} L.,  {Sargent} A.~I.,  1998, \apjl, 508, L91

\bibitem[\protect\citeauthoryear{{V{\'a}zquez-Semadeni}, {Passot} \&
  {Pouquet}}{{V{\'a}zquez-Semadeni} et~al.}{1995}]{Vazquez-Semadeni95}
{V{\'a}zquez-Semadeni} E.,  {Passot} T.,    {Pouquet} A.,  1995, in Revista
  Mexicana de Astronomia y Astrofisica Conference Series {MHD Turbulence, Cloud
  Formation, and Star Formation in the ISM (Invited paper)}.
pp 61--+

\bibitem[\protect\citeauthoryear{{Ward-Thompson}, {Andr{\'e}}, {Crutcher},
  {Johnstone}, {Onishi} \& {Wilson}}{{Ward-Thompson}
  et~al.}{2007}]{Ward-Thompson07}
{Ward-Thompson} D.,  {Andr{\'e}} P.,  {Crutcher} R.,  {Johnstone} D.,  {Onishi}
  T.,    {Wilson} C.,  2007, in {Reipurth} B.,  {Jewitt} D.,   {Keil} K.,  eds,
  Protostars and Planets V {An Observational Perspective of Low-Mass Dense
  Cores II: Evolution Toward the Initial Mass Function}.
pp 33--46

\bibitem[\protect\citeauthoryear{{Williams}, {Blitz} \& {McKee}}{{Williams}
  et~al.}{2000}]{Williams00}
{Williams} J.~P.,  {Blitz} L.,    {McKee} C.~F.,  2000, Protostars and Planets
  IV, pp 97--+

\bibitem[\protect\citeauthoryear{{Williams}, {de Geus} \& {Blitz}}{{Williams}
  et~al.}{1994}]{Williams94}
{Williams} J.~P.,  {de Geus} E.~J.,    {Blitz} L.,  1994, \apj, 428, 693

\end{thebibliography}

\label{lastpage}

\end{document}